\documentclass{style}
\usepackage{graphicx}
\usepackage{multirow}
\usepackage{rotating}
\usepackage{mathrsfs}
\usepackage{amsmath}
\usepackage{lscape}
\usepackage[usenames,dvipsnames]{color}
\usepackage{tikz,ifthen,calc}
\usetikzlibrary{snakes,shapes}
\usetikzlibrary{patterns}
\usetikzlibrary{plotmarks}
\usetikzlibrary{calc}
\usepackage{pgfplots}
\usepackage[utf8]{inputenc}
\usepackage{graphicx} 
\usepackage{amsmath,amssymb}
\usepackage[utf8]{inputenc}
\usepackage{graphicx} 
\usepackage{amsmath,amssymb}
\usepackage[square,sort,comma,numbers]{natbib}
\newcommand{\nn}{\nonumber\\}
\newcommand{\be}{\begin{equation}}
\newcommand{\ee}{\end{equation}}
\newcommand{\bea}{\begin{eqnarray}}
\newcommand{\eea}{\end{eqnarray}}
\newcommand{\bem}{\begin{multline}}
\newcommand{\eem}{\end{multline}}

\begin{document}

\title{$N=Z$ nuclei: A laboratory for neutron-proton collective mode}
\author{
 Chong Qi\email{chongq@kth.se}  and Ramon Wyss\\
  \it Department of Physics, Royal Institute of Technology (KTH), SE-10691 Stockholm, Sweden
}


\date{}

\maketitle

\begin{abstract}
The neutron-neutron and proton-proton pairing correlations have long been recognized to be the dominant many-body correlation beyond the nuclear mean field since the introduction of pairing mechanism by Bohr, Mottelson and Pines nearly 60 year ago. Nevertheless, few conclusion has been reached concerning the existence of analogous neutron-proton (np) pair correlated state. One can see a renaissance in np correlation studies in relation to the significant progress in radioactive ion beam facilities and detection techniques. The np pairs can couple isospin T = 1 (isovector) or 0 (isoscalar). In the isovector channel, the angular momentum zero component is expected to be the most importance one. On the other hand, as one may infer from the general properties of the np two-body interaction, in the isoscalar channel, both the np pairs with minimum (J=1) and maximum (J=2j) spin values can be important. In this contribution, we will discuss the possible evidence for np pair coupling from different perspective and analyze its influence on interesting phenomena including the 
Wigner effect  and mass correlations in odd-odd nuclei. In particular, we will explain the spin-aligned pair coupling scheme and quartet coupling involving pairs with maximum (J=2j) spin values.
\end{abstract}
\section{Introduction}
It is well known that, along the discovery of the uncharged particle neutron by Chadwick in 1932 \cite{Cha1932}, Heisenberg immediately introduced the idea that the nucleus is composed of protons and neutron as well as the concept of isospin \cite{Hei1932}. These mark the beginning of nuclear structure physics. The isospin  has been extensively applied in explaining many aspects of the nuclei \cite{tal93}, even though it is not a fundamental symmetry since the masses of the proton and neutron and the interactions involving the two are not exactly the same. Another glorious event is the suggestion of nuclear pairing mechanism by Bohr, Mottelson and Pines \cite{PhysRev.10.936} one year after the introduction of the Bardeen-Cooper-Schrieffer (BCS) ansatz in superconductors \cite{PhysRev.108.1175}. The neutron-neutron (nn) and proton-proton (pp) pairing correlations between identical particles have been shown to be crucial in explaining a wealth of experimental data has
including odd-even staggering in binding energies and charge radii, nuclear deformation as well as moments of inertia \cite{ring2004nuclear,bohr1998nuclear,bohr1998nuclear3}. It is natural to expect that the pairing correlation between the neutron and the proton can be equally important.
The neutron and proton can be coupled to both isoscalar $(T=0)$ pair, with a symmetric wave function for the radial spin part, and isovector $(T=1)$ pair in analogous to the nn and pp pair coupling.
However, there is no conclusive evidence for either np pair. 

There has been a long history and extensive efforts studying the importance of np correlations from many different perspectives. 
Recently, the np has attracted renewed interest in relation to the advances in experimental techniques and availability of radioactive beam facilities.  Overviews on np pairing correlations are published recently in Refs. \cite{Frauendorf201424,doi:10.1142/S0218301313300282,arXiv:1205.2134}. Extensive discussions can also be found in workshops organized recently (see, e.g., Ref. \cite{lee}). 
 The importance of np correlation in
the development of collective correlation  and nuclear deformation as well as in the evolution of
the shell structure has been generally accepted. This is related to the fact that the np two-body interaction contribute significantly to the nuclear mean field. The remaining controversy is whether it is necessary to include the residual np pairing coupling on top of the nuclear mean field and, if so, how to separate it from the mean field channel of the two-body interaction. Besides the dynamic effects of the residual correlation, another challenging task to understand the different predictions of the approximate methods and exact solutions within the shell-model context in treating the np correlation or the np coupling scheme. 
In the present contribution we will review briefly a few aspects   along that direction that are not fully covered in Ref. \cite{Frauendorf201424}. Then we will give a more detailed explanation on the works done at Stockholm during the past two decades.

\section{Systematics of nuclear binding energy and the residual correlations}

In a broader context, one may state that nuclear physics is an emergent phenomenon which is created by the complicated interplay among its constitutes: protons and neutrons. 
The understanding of its emergent behavior progresses by systematic experimental observations and the construction
of models to interpret them \cite{row10}, in particular their local fluctuations. 
Studies on the nuclear mass and other ground state properties reveal   strikingly systematic  behaviors including the nuclear liquid, shell structure well as the nuclear deformation.
It is thus natural to expect that the differences of binding energies can be used to isolate specific correlations. 
As mentioned above, the zigzag behavior of one-body separation energies and the nuclear binding energy has long been well known. It
provides clues to the pairing correlation between like nucleons. It may also be possible to extract the residual interactions between protons and neutrons from the binding energy differences.

\subsection{Like-particle pairing and odd-even staggering}
The pairing energy or pairing gap implies that the energies of even-even nuclei are systematically lower than those of odd-odd and odd-$A$ nuclei. 
As commented in P. 169 of Ref. \cite{bohr1998nuclear}, one can extract the empirical pairing gap by comparing the local average of the masses of odd-$A$ nuclei with the masses of the corresponding even-even nuclei, where one assumes that
the masses are a smooth function of $Z$ and $N$ except for the pairing effect.

The four-point formula used to extract the neutron empirical pairing gap from the odd-even staggering (OES) of the binding energies is defined as~\cite{bohr1998nuclear}
\begin{multline}
\label{eq:4point}
\Delta^ {(4)}(N)= \frac{1}{4} [-B(N+1,Z)+3B(N,Z)\\
-3B(N-1,Z)+B(N-2,Z)],
\end{multline} 
where $B$ is the (positive) binding energy and $N$ and $Z$ are assumed to be even numbers.  The proton pairing gap can be defined in a similar way. 
The trend of the extracted pairing gaps can be well approximated as
$\Delta\approx 12/A^{1/2}$ MeV. However, as already noted in Ref. \cite{bohr1998nuclear}, above filter shows large local fluctuations and appears to be correlated with the
shell structure.

The three-point formula is a simpler expression one can use to extract the empirical pairing gap from the 
binding energy
\cite{bohr1998nuclear,PhysRevLett.81.3599}, which, for systems with even 
neutrons, has the form~\cite{PhysRevLett.81.3599}
\begin{multline}
\label{eq:3point}
\Delta^ {(3)}(N)=-\frac{1}{2}\left[B(N-1,Z)+B(N+1,Z)-2B(N,Z)\right]\\
=-\frac{1}{2}[S_n(N+1,Z)-S_n(N,Z)]
\end{multline} 
where $S_n$ is the one-neutron 
separation energy.
Above formula indicates that $\Delta^ {(3)}(N)$ measures the additional 
binding gain by the last neutron in the even-$N$ system relative to the odd 
system with one more neutron. 
It should be mentioned that, besides pairing, a number of other mechanisms may contribute to the OES 
\cite{PhysRevLett.81.3599,PhysRevC.88.064329,Friedman2009,PhysRevC.60.051301}. This includes
effects induced by the mean field in deformed nuclei (or the Kramers 
degeneracy) and the contribution from the diagonal
interaction matrix elements of the two-body force. 
As discussed in detail in Refs. 
\cite{PhysRevLett.81.3599,PhysRevC.63.024308},
in even systems where the last neutrons occupy different orbitals the single-particle energy
contributes substantially to $\Delta^ {(3)}(N)$.

\begin{figure}
\includegraphics[width=0.45\textwidth]{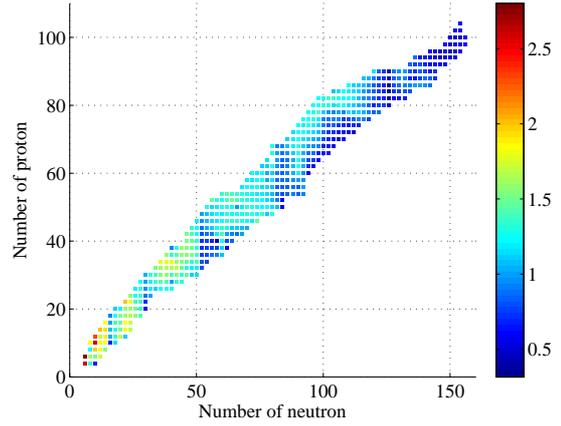}
\caption{\label{fig:4Gaps_errors} Neutron pairing gaps (in MeV) calculated by Eq. (\ref{eq:3pointC}) for all even-even nuclei}
\end{figure}

The contribution from the quickly varying single-particle structure of the mean field to the empirical  pairing gap is minimized in odd systems.
One can propose another version of the three-point formula as
\begin{multline}
\label{eq:3pointC}
\Delta^ {(3)}_{C}(N)=\frac{1}{2}[S_n(N,Z)-S_n(N-1,Z)]\\
=\frac{1}{2}\left[B(N,Z)+B(N-2,Z)-2B(N-1,Z)\right] \\
=\frac{1}{2}[S_{2n}(N,Z)-2S_n(N-1,Z)],
\end{multline} 
which actually corresponds to $\Delta^ {(3)}$ for the case of odd nuclei 
\cite{PhysRevLett.81.3599}. The values of $\Delta^ {(3)}$ extracted from experimental binding energies \cite{1674-1137-36-12-003,Audi03} are plotted in Fig. \ref{fig:4Gaps_errors}. One may state that $\Delta^ {(3)}_C(N)$ measures the pairing effect in the odd nuclei, whereas $\Delta^ {(3)}(N)$ is impacted by single particle states. The value of $\Delta^ {(3)}(N-1)$ has often been compared to the theoretical pairing gap calculated for the even systems \cite{Lesinski2009,PhysRevC.89.054320} and to the OES derived from theoretical binding energies \cite{PhysRevC.85.014321,PhysRevC.79.034306}. The direct comparison between the theoretical pairing gap and empirical OES is convenient from a computational point of view since only one single calculation is required, which avoids the complicated handling of the blocking effect in the odd nuclei.

The two three-point formulas are often written together as
\begin{multline}
\label{eq:3pointC}
\Delta^ {(3)}(N)=\frac{(-1)^{N+1}}{2}\\
\times\left[B(N-1,Z)+B(N+1,Z)-2B(N,Z)\right] 
\end{multline}
where $N$ takes both even and odd numbers and  $(-1)^{N+1}/2$ defines the number parity. However, it should be emphasized that, as we understand now, the physics behind the two quantities are very different.

 The four-point formula can be rewritten as
\begin{multline}
\label{eq:4point2}
\Delta^ {(4)}(N)
=\frac{1}{2}[\Delta^{(3)}(N)+\Delta^{(3)}_C(N)].
\end{multline} 
That is, it measures the average value of $\Delta^ {(3)}$ in adjacent even and odd systems.

There are other formulas available for the pairing gap.
The five-point formula is given by~\cite{NuclPhysA.476.1,NuclPhysA.536.20,PhysRevC.65.014311}
\begin{multline}
\label{eq:5point}
\Delta^ {(5)}(N)= \frac{1}{8} [B(N+2,Z)-4B(N+1,Z)\\
+6B(N,Z)-4B(N-1,Z)+B(N-2,Z)]\\
=\frac{1}{4}[\Delta^{(3)}_C(N+2)+2\Delta^{(3)}(N)+\Delta^{(3)}_C(N)].
\end{multline}
It indicates that the shell effect is still present in $\Delta^{(5)}$.
The five-point formula is also used in Refs.
\cite{PhysRevC.60.051301,PhysRevC.87.064308,Bender2000}. $\Delta^{(4)}(N)$ and $\Delta^{(5)}(N)$ show quite similar results for most nuclei since $\Delta^{(3)}_C(N)$ varies smoothly \cite{Changizi2015210,PhysRevC.91.024305}. 
In Refs. \cite{PhysRevC.88.034314,Dobaczewski01032002}, the experimental pairing gap is taken as the average of adjacent ones deduced through the three-point formula as
\begin{eqnarray}
\begin{aligned}
\Delta_{ave}^{(3)}(N)=\frac{1}{2}\left[\Delta^{(3)}_{C}(N) + \Delta^{(3)}_{C}(N+2)\right],
\end{aligned}
\end{eqnarray}
which is actually also a five-point formula involving the same group of nuclei as $\Delta^ {(5)}(N)$ but with different weights for each nucleus. 
Again there is no significant 
difference between the results derived from $\Delta_{ave}^{(3)}(N)$ and 
$\Delta^{(3)}_{C}(N)$ for open-shell nuclei where the pairing gap is a smooth 
function of $N$. Noticeable differences between $\Delta^{(3)}_{ave}(N)$ and $\Delta^{(3)}_{C}(N)$ 
may be seen where abrupt changes in pairing correlations are expected to 
happen, e.g., around shell closures, which is smoothed out in the former case.  A quite sophisticated version of $\Delta^{(3)} (N)$ is used in Ref. \cite{PhysRevC.88.064329} by subtracting the liquid-drop and shell effect contributions to the binding energy. The results are similar to those of $\Delta^{(3)}_C (N)$. In particular, they show a quite similar isospin dependence.

$\Delta^ {(3)}_C(N)$ show a much weaker $A$ dependence than other formulae. Actually it can be reasonably fitted as a constant value. This agrees with the suggestion in Ref. \cite{Friedman2009} that the pairing gap may not show any $A$ dependence. In \cite{bohr1998nuclear} it is commented that the pairing energy derived from Eq. (\ref{eq:3point}) is systematically too small. However, this is definitely not the case for $\Delta^ {(3)}(N)$ which are systematically larger than all the other three cases.
$\Delta^ {(3)}_C(N)$ show the smallest values due to the reason that they largely remove the contribution from the mean field. 
In fact, the differences between $\Delta^ {(3)}(N)$ and $\Delta^ {(3)}_{C}(N)$ largely reflects the gap between the corresponding neighboring orbitals \cite{PhysRevLett.81.3599}. Another essential difference between $\Delta^{(3)}_C (N)$ and all other mentioned OES formulas is that $\Delta^{(3)}_C (N)$ diminish for closed-shell nuclei, which is in agreement with our common expectation that the pairing effect diminish at shell closure due to the reduced level density.

\begin{figure}
\includegraphics[width=0.45\textwidth]{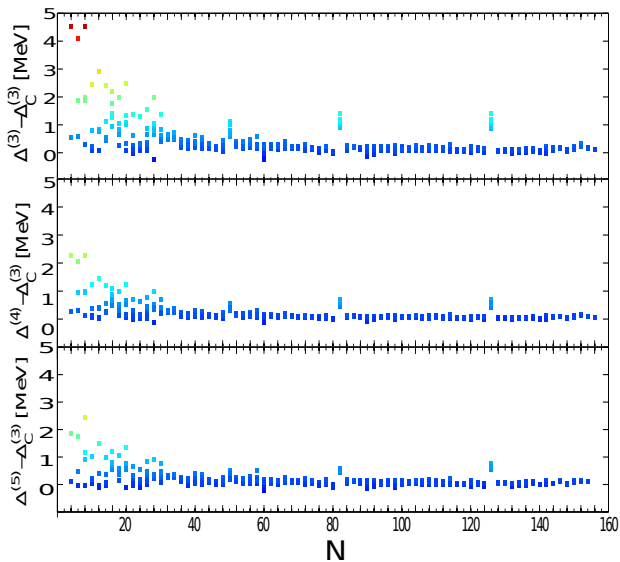}
\caption{\label{fig:Diff_Gaps_subplot3} Differences between different gap formulae with respect to $\Delta^{(3)}_{C}$.}
\end{figure}

$\Delta^ {(3)}_C(N)$ is smaller than $\Delta^ {(3)}(N)$, $\Delta^ {(4)}(N)$ as well as $\Delta^ {(5)}(N)$ in most cases. The differences between the various gap formulae and the 3-point formula $\Delta^{(3)}_{C}$ are plotted in Fig. \ref{fig:Diff_Gaps_subplot3}. The dispersal of the data below $N<30$ is apparent. This can be an indication of the significant mean-field contribution to the three formulae in this region, which is expected to show a $A^{-1}$ dependence \cite{Friedman2009}.

\begin{figure}
\includegraphics[width=0.45\textwidth]{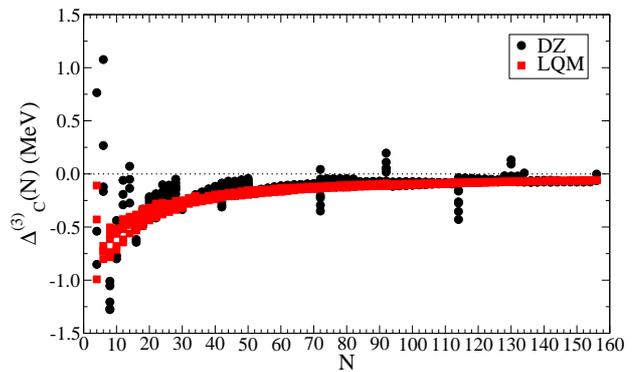}
\caption{ Contributions to $\Delta^{(3)}_{C}$ from the liquid drop model (LQM) and Duflo-Zuker (DZ) shell-model mass formulas. }\label{fig3}
\end{figure}

In the desired case, if the even nucleus of concern and the intermediate odd nucleus show the same mean field property, $\Delta^{(3)}_{C}$ is not expected to contain any contribution from the mean field \cite{PhysRevLett.81.3599}. 
However, as pointed out by Bohr and Mottelson  in P. 171 of Ref. \cite{bohr1998nuclear}, the average binding energy contains significant terms that are not linear in $N$.
As a result, from a macroscopic point of view, there may be a residual contribution to $\Delta^{(3)}_{C}$ from the symmetry energy (expected to be negative) and other non-linear terms. To illustrate this point, in Fig. \ref{fig3} we evaluated the $\Delta^{(3)}_{C}$ values from the liquid drop model (LQM) with no shell or pairing energy corrections and the Duflo-Zuker (DZ) shell model mass formula \cite{Duf95} by removing the pairing term. The parameters of the LQM and DZ models are taken from Ref. \cite{PhysRevC.92.024306} and \cite{0954-3899-42-4-045104}, respectively. Calculations with the DZ model show more local fluctuations than those of the LQM, which has not been understood yet. But both cases show a vanishing behavior as $A$ increases. 
For heavy nuclei, this residual contribution is around -60 keV. This amount is acceptable by taking into account the fact the in practice the OES is a result of the delicate interplay between the mean field and the pairing correlation.
It is hoped that, by using $\Delta^{(5)}$, the smooth non-linear terms in the binding energy can be canceled up to the fourth order. However, as mentioned in Ref. \cite{PhysRevC.88.064329}, the odd-even effects may be diminished as a result of the averaging over nuclei further apart.

$\Delta^{(3)}_{C}(N)$ contains fruitful information on the pairing effects. It is the best to remove the contribution from the varying part of the nuclear 
mean field as well as contributions from other shell structure details and can serve as a good measure of the pairing effect in in even-$N$ systems. Moreover, by using $\Delta^{(3)}_{C}(N)$ one can make it more convenient to extract the neutron-proton interaction from binding energy differences \cite{Qi2012436}.

The possible isospin dependence of the empirical odd-even staggering was discussed in Ref. \cite{PhysRevLett.95.042501}.

Within the shell model context, one has to separate the contribution from the monopole channel of the two-body interaction when studying the residual two-body correlations. For simple systems within a single-$j$ shell with a monopole pairing coupling $G$, the total energy of a system with $n$ particles can be written as \cite{tal93,Qi2012436}
\begin{eqnarray}
E&=&\varepsilon n + \frac{2a-G}{4}n(n-1)\\
\nonumber&&+\frac{b-2G}{2}\left[\mathcal{T}(\mathcal{T}+1)-\frac{3n}{4}\right]\\
\nonumber&&+(j+1)G(n-v)+G\left[\frac{v^2}{4}-v+s(s+1)\right],
\end{eqnarray}
where $\varepsilon$ is the single-particle energy, $a$ and $b$ defines the monopole interactions for the orbital $j$, $v$ is the seniority quantum number, $T$ is the total spin, $s$ the reduced isospin. The monopole interaction is defined as the angular momentum weighted average of all two-body interaction matrix elements. They are independent of angular momentum and their contribution to the energy (or diagonal matrix element of the Hamiltonian matrix) are only related simply to the particle number and total isospin.
The reduced isospin is related to the isospin of the states with seniority $v$ in the $j^v$ configuration, from which one can realize that we always have $t\leq v/2$. For systems with the maximum isospin ($T=n/2$ as in above case), the reduced isospin of any seniority $v$ state is $t=v/2$. The (ground) state with $v = 0$ is uniquely defined
with reduced isospin $t = 0$ for any $j^n$ configuration. In addition, the
$v = 1$ state is a state with $J = j$ and $t = 1/2$. 
 From above equation it can be seen that it is the term $(j+1)Gv$ that may result in an odd-even staggering in nuclear binding energies. This suggested that the residual pairing term in macroscopic mass formulas may be written as
\begin{equation}\label{pt}
E_p\propto 2-v,
\end{equation}
where $v=1$ for odd-$A$ nuclei and $v=2$ for the $\mathcal{T}=|N-Z|$/2 ground state of odd-odd nuclei. There should be no additional gain in pairing energy when crossing the $N=Z$ line.

\subsection{The empirical neutron-proton interaction}

The (phenomenological) average interaction between the
last protons and the last neutrons in even-even
nuclei can be extracted from the double difference of binding
energies as~\cite{Zhang19891}
\begin{eqnarray}\label{vpn-ee}
\nonumber V_{ee} {(Z, N)}&=& \frac{1}{4}\left[
B(Z,N)+B(Z-2,N-2)\right. \\ 
&&- \left.B(Z-2,N)-B(Z,N-2) \right],
\end{eqnarray}
where $B(Z,N)$ is the (positive) binding energy of a nucleus with $Z$ protons and $N$ neutrons. 
The factor $1/4$ takes into account the fact that four additional pairs are 
formed by the last two protons and neutrons. 
 $V_{pn}$ extracted from experimental nuclear binding energies \cite{Audi03}  evolve rather smoothly as a function of mass number $A$. 
In fact, this average behavior of $V_{pn}$ also probes the symmetry energy term (i.e., the isospin-dependence 
of the binding energy) in the macroscopic mass formula. The overall trend of $V_{pn}$ can be 
well approximated by a smooth relation of $(a+a_sA^{-1/3})/A$ \cite{Sto07,Qi2012436}.
Above formula has been extensively analyzed in Refs. \cite{Cakirli2005,Cakirli2006,Oktem2006,Neidherr2009a,Cakirli2009,Brenner2006,Chen2009a}.

\begin{figure}[htdp]
\includegraphics[width=0.45\textwidth]{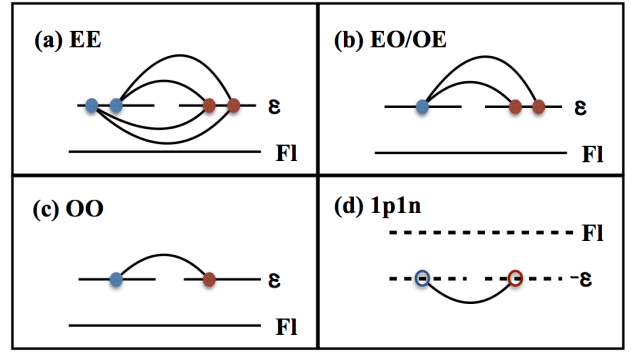}
\caption{\label{shiyi} (Color online) Illustration of the np interaction in even-even (a), odd-$A$ (b) and odd-odd (c) nuclei as extracted from Eqs. (\ref{vpn-ee}-\ref{vpn-oo}) as well as those from Eq. (\ref{vpn-zhao}) for even-even nuclei, which corresponds to a hole-hole-like np interaction. }
\end{figure}

The average np interaction in odd-$A$ and odd-odd
nuclei can be extracted in a similar way as
\begin{eqnarray}\label{vpn-eo}
 && V_{eo} {(Z, N-1)}= \frac{1}{2}[
B(Z,N-1)+B(Z-2,N-2) \hspace{5mm}\nonumber\\
&&\hspace{2.2cm}-B(Z-2,N-1)-B(Z,N-2)],
\end{eqnarray}
\begin{eqnarray}\label{vpn-oe}
\nonumber && V_{oe} {(Z-1, N)}= \frac{1}{2}[
B(Z-1,N)+B(Z-2,N-2)\hspace{5mm} \\
&&\hspace{2.2cm}-B(Z-2,N)-B(Z-1,N-2)],
\end{eqnarray}
which involve two np pairs and
\begin{eqnarray}\label{vpn-oo}
&&V_{oo} {(Z-1, N-1)}=\hspace{5cm}\nonumber\\
&&\hspace{1.8cm}\nonumber B(Z-1,N-1)+B(Z-2,N-2)\\
&&\hspace{1.7cm}-B(Z-1,N-2)-B(Z-2,N-1),
\end{eqnarray}
involving one np pair. We have assumed that $N$ and $Z$ only take even numbers in above equations.

In addition to the family of mass relations shown above, there is another way to extract the average np interaction as 
\begin{eqnarray}\label{vpn-zhao}
  V_{1n-1p} {(Z, N)}= B(Z,N)+B(Z-1,N-1)  \nonumber \\
- B(Z,N-1)-B(Z-1,N), 
\end{eqnarray}
which is irrespective to the oddness of the proton and neutron numbers. This was proposed in Ref. \cite{Bas71}
and applied recently in Refs. \cite{zhao15,zhao11,zhao13,Zhao14}. This equation is identical to Eq. (\ref{vpn-oo}) studied above for the cases of odd-odd nuclei, but involves, in the other cases, the breaking of the proton and/or neutron pairs for which one intentionally avoid in the construction of the first family of average np interaction.
By comparing Eqs. (\ref{vpn-oo}) and (\ref{vpn-zhao}) one interesting thing we notice is that one can re-interpret $V_{1n-1p}$ for even-even nuclei as a measure of the np interaction between the two neutron and proton holes relative to the even-even 'core' nucleus. This is illustrated in the lower-right panel of Fig. \ref{shiyi}. Indeed, the values for $V_{1n-1p}$ for even-even nuclei follow roughly the same trend as those for the odd-odd nuclei with one less np pair. The latter case coincide with Eq. (\ref{vpn-oo}). This is also related to the fact that, as mentioned in Ref. \cite{zhao15}, the average np interactions for even-even and odd-odd nuclei as extracted from the binding energies predicted by the Garvey-Kelson mass relations \cite{PhysRevLett.16.197} satisfy 
\begin{eqnarray}\label{vpn-GKs}
 V_{1n-1p} {(Z, N)}- V_{1n-1p} {(Z+1, N+1)}&\cong& 0.
\end{eqnarray}
One more thing that one should notice is that both above values are larger than $V_{ee}$ for the neighboring even-even nuclei. The reason will be analyzed in the next section.

Stoitsov {\it et al.} showed that the global 
properties of $V_{pn}$ can be reproduced by 
Hartree-Fock-Bogoliubov (HFB) calculations with the Skyrme functional plus a density-dependent $\delta$ pairing 
interaction~\cite{Sto07}. A detailed calculation was also done in Ref. \cite{Ben11} where the effects of the deformation and collective fluctuation on $V_{pn}$ were analyzed. It is still quite interesting
to explore the local fluctuations of $V_{pn}$ around the average values which large-scale HFB calculations 
fail to explain~\cite{Sto07}, which may carry further
nuclear structure information and serve as a constraint in future developments of nuclear structure models.

It is understood that $V_{pn}$ also measures the
extra binding gained by the neutron (proton) pair when two additional protons (neutrons) are added.
The two-nucleon separation energies in even-even nuclei can be written as
\begin{eqnarray}
S_{2n}(Z,N)=2S_n(Z,N-1)+2\Delta^{(3)}_C(Z,N).
\end{eqnarray}
 Eq. (\ref{vpn-ee}) can be rewritten as
\begin{eqnarray}
\nonumber V_{ee} {(Z, N)}&=&\frac{1}{2}\left[S_{n}(Z,N-1)-S_{n}(Z-2,N-1)\right]\\
&&+\frac{1}{2}\left[\Delta^{(3)}_C(Z,N)-\Delta^{(3)}_C(Z-2,N)\right].
\end{eqnarray}
The quantities 
$\delta S_n(Z,N-1)=S_{n}(Z,N-1)-S_{n}(Z-2,N-1)$ and $\delta_n(Z,N)= \Delta_C^{(3)}(Z,N)-\Delta_C^{(3)}(Z,N-2)$
measure the isospin dependences of the one-body separation energy (the mean-field) and pairing 
interaction, respectively. One can easily see that $\delta S_n(Z,N-1)$ (and $\delta S_p(Z-1,N)$) also
reveals the average proton-neutron interaction between the last proton pair and odd neutron as
\begin{eqnarray}\label{vpn-o}
 V_{eo}(Z,N-1)&=&\frac{1}{2}\delta S_n(Z,N-1).
\end{eqnarray}

Contributions from the two basic ingredients $\delta S$ and $\delta$ on the empirical proton-neutron 
interaction $V_{pn}$ can be extracted from experimental nuclear masses.
It is seen that $V_{pn}$ is dominated by the contribution from $\delta S$. The $\delta_n$ and $\delta_p$ values are 
comparatively small, mostly within $|\delta|\leq 100$~keV. This indicates that the empirical 
proton-neutron interaction
can to a large extent be understood as a mean-field or symmetry energy effect. It is also consistent with the observation of Ref.~\cite{Sto07}
that HFB calculations on $V_{pn}$ are 
insensitive to the different choices of pairing forces.

Empirical studies of the nuclear masses suggest that the average np interactions for even-even and neighboring odd-$A$ nuclei thus extracted from experimental data are roughly the same and show a rather smooth behavior as a function of $A$ in most cases \cite{Qi2012436}. 
If the local fluctuations in the pairing interactions are negligible, it should be
\begin{equation}\label{e-o}
V_{pn}(Z,N)\approx V_{pn}(Z,N-1)\approx V_{pn}(Z-1,N),
\end{equation}
whereas those for the odd-odd nuclei are systematically larger than the former ones. 

 For a $I=j$, $T=1/2$ system with three particles in a single-$j$ shell, we have $v=1$ and $s=1/2$. The $V_{pn}$ for such a nucleus can be expressed in the same form as above. The empirical relation of Eq.~(\ref{e-o}) still holds for these self-conjugate nuclei. In reality we have
\begin{equation}
V_{pn}(Z,Z)\approx V_{pn}(Z,Z-1)\approx V_{pn}(Z-1,Z),
\end{equation} 
where $Z$ takes even values.
 
The empirical interactions between the odd proton and odd neutron in odd-odd nuclei can be extracted from binding energies
in a way similar to those of even-even and odd-$A$ systems. The ground state of odd-odd $N=Z$ nuclei may carry isospin quantum numbers $T=0$ or 1. For the lowest $T=0$ state one may extract the proton-neutron interaction as
\begin{eqnarray}\label{pno}
\nonumber &&V_{pn}(Z-1,Z-1) \\
\nonumber &&= B(Z-1,Z-1)+B(Z-2,Z-2)\\
\nonumber&& -B(Z-1,Z-2)-B(Z-2,Z-1)\\
&&=\frac{3b}{4}-a.
\end{eqnarray}
It indicates that the $V_{pn}$ in odd-odd $N=Z$ nuclei are three times as large as the average values in $N\neq Z$ nuclei while those in even-even $N=Z$ nuclei and the adjacent odd-$A$ nuclei with one less nucleon are roughly twice as large as those in neighboring $N\neq Z$ nuclei. In reality $b$ should be positive. In medium mass and heavy nuclei, it should also be much larger than the pairing strength $G$. 
In the spin-isospin SU(4) symmetry limit, the $V_{pn}$ of $N=Z$ nuclei are four times larger than those for $N\neq Z$ \cite{PhysRevLett.74.4607}. There was also no difference between $V_{pn}$ in even-even and odd-odd $N=Z$ nuclei.

\subsection{Residual neutron-proton interaction in odd-odd nuclei}
On P. 171 of Ref. \cite{bohr1998nuclear}, Bohr and Mottelson pointed out that  there is a systematic tendency for an extra binding of the odd-odd nuclei. It may be a result from the residual interaction between the last unpaired neutron and the unpaired proton in those nuclei.
The average np interaction $V_{oo}$ in Eq. (\ref{vpn-oo}) were often compared with the $T=1$ proton-proton and neutron-neutron pairing interactions. In reality, $V_{oo}$ is a mixture of the mean field effect and the re-coupling effect due to the residual np interaction between the two unpaired particles. The mean field effect has to be properly filtered out if one aims at studying the residual np coupling. This is important for our eventual clarification of the role played by np pairing correlation in nuclei.

From a phenomenological point of view, it is understood that $V_{ee}$, $V_{eo}$ and $V_{oe}$ are dominated by contributions from the nuclear symmetry energy which is induced by the monopole np interaction.
The non-vanishing values for $\delta_{np}$ as extracted from experimental data can be a measure of the residual/additional np re-coupling energy, which can be rewritten as
\begin{eqnarray}\label{vpn-sc2}
\delta_{np} &=&\Delta^{(3)}_{n} (Z,N) -\Delta_n^{(3)} (Z-1,N) \nonumber \\
&=& \frac{1}{2}\left[
B(Z,N)+B(Z,N-2) \right. \nonumber \\
&&- B(Z-1,N)-B(Z-1,N-2)  \nonumber \\
&&- \left. 2B(Z,N-1)+2B(Z-1,N-1) \right]\nonumber \\
&=&V_{oo}-[2V_{eo}+V_{oe}-2V_{ee}]\nonumber\\
&=&\frac{1}{2}[V_{1n-1p}(Z,N)-V_{1n-1p}(Z,N-1)].
\end{eqnarray}
That is, it corresponds to the difference between $V_{oo}$ and an weighted average of those for odd-$A$ and even-even nuclei or half the difference between $V_{1n-1p}$ for even-even nuclei and odd-$A$ nuclei. A very similar result can be obtained by taking the difference between the pairing gaps of the even-even and corresponding even-$Z$-odd-$N$ nucleus.
There are also different ways to extract the residual np interaction \cite{Friedman2007, Jensen1984393,Wu15a,NuclPhysA.476.1,NuclPhysA.536.20}.

\begin{figure}
\includegraphics[width=0.45\textwidth]{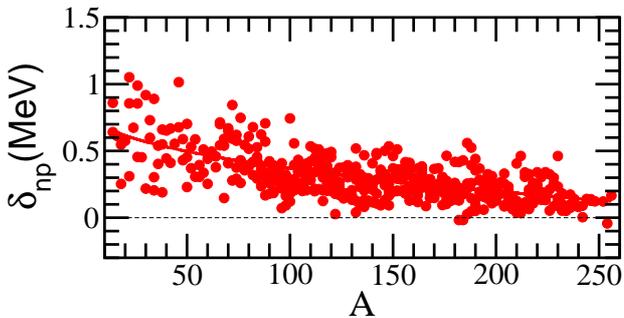}
\caption{\label{csexp} (Color online) The residual np interactions defined in Eq. (\ref{vpn-sc2}) for all known nuclei as extracted from experimental data.}
\end{figure}

The residual np interaction as extracted from experimental binding energies \cite{1674-1137-36-12-003} by using above three formulas are plotted in Fig. \ref{csexp}. It is seen that, as expected, the extracted $\delta_{np}$ values are positive for almost all known nuclei. The values of all above three formulas show a weak dependence on the mass number $A$ with large fluctuations. The mean $\delta_{np}$ values are around 300 keV in all cases. The values for available superheavy nuclei reduces to below 200 keV.

The positive contribution from the residual np interaction to the total binding energy is the origin of the odd-even staggering in $V_{1n-1p}$ that was studied in Ref. \cite{zhao15}.

The additional binding in odd-odd nuclei is due to the residual interaction between the odd proton and neutron. That enhancement can be extracted from binding energy differences and is beyond the mean-field treatment. The challenge, however, is how to differentiate between a correlated np pair (in the BCS sense) and those uncorrelated ones. 

\subsection{The Wigner energy}
Wigner noticed that there are large changes in binding energy in nuclei with approximately the same numbers of neutrons and protons \cite{Wigner41}. This has often been referred to as the Wigner effect.  

The residual np interaction $V_{ee} {(Z, N)}$ discussed above have been applied in the study of the Wigner energy. In Ref. \cite{PhysRevLett.74.4607}, van Isacker, Warner, and Brenner showed that the large double binding energy differences for even-even $N=Z$ nuclei can be a consequence of Wigner's SU(4) symmetry. Ref. \cite{Satula1997a} studied the separated contributions of neutron-proton pairs of a given angular momentum and isospin to the Wigner energy. It is also suggested that the Wigner term can be traced back to the isospin $T = 0$ part of nuclear interaction. It cannot be solely explained in terms of correlations between the neutron-proton $J = 1$, $T = 0$ pairs. 

In Ref. \cite{PhysRevLett.99.082501} it is argued that
the discrepancies between empirical shell gaps determined from binding energy differences and the gaps calculated with mean-field models can be resolved by taking into account the Wigner energy in the even-even $N=Z$ nuclei and the corresponding nuclei with one less nucleon as induced by the diagonal correlation energy due to nn, pp, and np pairing interactions. The large difference between observation and mean-field calculation was already noticed by Bohr and Mottelson (see, Fig. 3-5 and P. 328 of Ref. \cite{bohr1998nuclear}). It was pointed out that the calculated spectra reproduce approximately the observed positions of the single-particle levels above the Fermi surface, but underestimate the binding of deep-lying hole states. 
The increased binding of these states may be interpreted in terms of a velocity dependence of the mean field \cite{bohr1998nuclear}. 

There are still extensive efforts trying to determine the Wigner energy in a precise way.
One thing has to be considered is the effect of the symmetry energy on the extraction of the Wigner effect. The extracted Wigner energy can be very different if one takes the symmetry energy of the from $T(T+1)$ instead of the normal $(N-Z)^2$.

\section{Seniority and np coupling schemes}
The low-lying yrast states in $^{92}_{46}$Pd were recently 
reported  \cite{ced11}. This is the
heaviest $N=Z$ nucleus with measured  spectrum so far. It was 
suggested that in this nucleus, as well as in neighboring nuclei like 
$^{96}$Cd, the properties of the low-lying states can be 
classified by a spin-aligned $np$ pair coupling scheme \cite{ced11, qi11}. 
That is, the ground state wave functions do not consist 
mainly of pairs of neutrons ($\nu\nu$) and protons 
($\pi\pi$) coupled to zero angular momenta, but rather of isoscalar 
$np$ pairs ($\nu\pi$) coupled to the maximum angular momentum 
$J$, which in the shell $0g_{9/2}$ is $J=9$   \cite{ced11, qi11}. 
A detailed shell-model analysis of the spin-aligned $np$ pair coupling was 
performed in Refs. \cite{qi11,Xu2012} based on coefficients of fractional 
parentage and multistep shell model calculations.

\begin{figure}[htdp]
\includegraphics[scale=0.35]{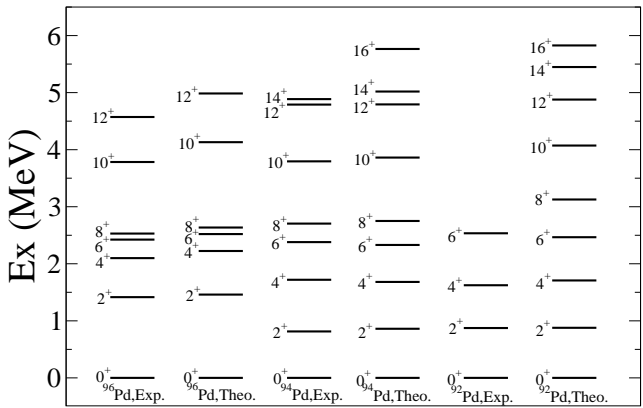}\\
\caption{
The observed positive parity yrast states of $^{92,94,96}$Pd and shell-model calculations from Ref. \cite{qi11}.}
\label{le9094}
\end{figure}

In Fig.~\ref{le9094} 
we plotted the experimental spectra of three even-$A$ neutron-deficient 
Pd isotopes.
As one would expect, in
$^{96}$Pd with four proton holes with respect to  $^{100}$Sn, 
the positions of the energy levels correspond to a $(g_{9/2})_\lambda ^2$ isovector
pairing (or seniority) spectrum.  In particular, the yrast $8^+$ state, which has the largest spin among the $v=2$ states, shows a typical pattern for isomeric states as induced by the seniority coupling. When the number of neutron holes increases, approaching $N=Z$, the levels tend to be equally separated. The lowest excited states in $^{92}$Pd exhibit a particularly regular pattern. The regular pattern was first noticed by J. Blomqvist. He explained that in terms of aligned np pairs in a scheme that is similar to the stretched scheme as described in Ref. \cite{PhysRev.161.1034}. He also expected that systems with such couplings should be deformed. 

In the following we will briefly describe the formalism used in above papers \cite{qi11,Xu2012} and in Refs. \cite{Qi2012d,
Qi2012b,Qi2012e,1742-6596-338-1-012027}. We will construct the basis in the neutron-proton 
representation. We will show that it is a quite natural extension of the seniority coupling within the shell-model context. 
There are quite many recent publications following the same direction or re-interpret the results in slight different ways (see, Refs. \cite{Zer11,PhysRevC.91.064318,PhysRevC.91.054323,PhysRevC.90.014318,PhysRevC.89.014316,PhysRevC.88.034329} and references contained therein).
A detailed investigation on the applications of the nucleon-pair approximation of the shell model on this subject is presented in Ref. \cite{Zhao20141}.
 In any case the isospin symmetry is exactly
conserved, as it is included in the 
interaction matrix elements that we use \cite{qi11}.

\subsection{The seniority coupling}
Many features in nuclear structure physics can be understood in term of the 
seniority coupling scheme. It showed to be extremely useful for the 
classification of nuclear states in the $jj$-scheme 
\cite{tal93}, particularly in semi-magic nuclei with only 
one type of nucleons. The driving force behind the dominance of seniority coupling is the strong pairing interaction between like particles. The study of the seniority coupling in single-$j$ systems is still an very active field \cite{Escuderos2006a,Qi2011a,Qi2010,Qi2012a,ISI:000304404400008}, in particular in relation to the possible existence of partial dynamic symmetry.

The seniority coupling scheme can be generalized to systems within many shells. The diagonalization of the Hamiltonian matrix within the seniority $v=0$ model space provides a way to solve the pairing Hamiltonian in an exact way \cite{PhysRevC.89.014321,Zelevinsky2003,Volya2001,Xu2013247,Changizi2015210}.
It is known that there is one nontrivial solution from the BCS ansatz which is interpreted as the nuclear ground state. Within the exact pairing scheme, there are as many eigenstates as the number of seniority $v=0$ states as defined within the model space. The state with the lowest energy is usually linked to the nuclear ground state. But it can still be interesting to understand the fundamental differences between the lowest-lying states and the excited states. 
 
\subsection{Number of pairs}
An interesting quantity is the so-called number of pairs or number of interaction links which can give a simple idea on the relative importance of different two-body interactions and pairs on the total energy as well as on the nuclear wave function. 
For nucleons in a single-$j$ shell, the correlation energy can be written as \cite{Qi10}
\begin{equation}
E_I=C_J^IV_J,
\end{equation}
where $I$ is the total angular momentum, $V_J=\langle j^2;J|\hat{V}|j^2;J\rangle$ are two-body matrix elements and $C^I_J$ is the number of pairs with angular momentum $J$. If isospin symmetry is conserved in the two-body interaction $\hat{V}$, then one has a total number of $2j+1$ matrix elements with angular momenta $J=0$ to $2j$.
The total number of nucleon pairs is given by \cite{DeGuerra2003,Qi10},
\begin{equation}
\sum_{J}C^I_J=n(n-1)/2,
\end{equation}
and
\begin{equation}
\sum_{J,{\rm odd}}C^I_J=\frac{1}{2}\left[\frac{n}{2}\left(\frac{n}{2}+1\right)-T(T+1)\right],
\end{equation}
where $n$ is the total number of nucleons and $T$ is the total isospin quantum number.

For a fully-filled single-$j$ shell, one has $E_0=\sum_{J}(2J+1)V_J$. This is simply related to the contribution from the monopole interaction and there is no correlation. The monopole interaction is defined as the weighted average of the two-body interaction matrix elements as
\begin{eqnarray}
\nonumber V_{jj'}&=&\frac{\sum_J (2J+1) V^{J}_{jj'jj'}}{\sum_J (2J+1) (1-\delta_{jj'}(-1)^{J})}\\
&=&\frac{\sum_J (2J+1) V^{J}_{jj'jj'}}{(2j+1)}\frac{1+\delta_{jj'}}{2j'+1-\delta_{jj'}}.
\end{eqnarray}
Their contribution to the total energy corresponds to
\begin{eqnarray}
E_m=\sum_{jj'}V_{jj'}\left\langle\frac{ n_j(n_{j'}-\delta_{jj'})}{1+\delta_{jj'}}\right\rangle,
\end{eqnarray}
where $n$ denotes the number of particles instead of pairs.
If only the pairing interaction is considered for the particle-particle channel, we have
$V^{J=0}_{jjjj}=-\Omega_jG_{jj}$ and
$V_{jj}=-G_{jj}/2j$.

\subsection{Simple systems with two np pairs}
The neutron-proton ($np$) correlation breaks the seniority symmetry in a major way.
Correspondingly, the wave function is a mixture of many components with different seniority quantum numbers. It is not clear yet how this kind of states can be classified in the $jj$-scheme. The stretch scheme, which corresponds to the maximally aligned intrinsic angular momentum, was proposed in the 1960s to describe the rotational-like spectra of open-shell nuclei \cite{PhysRev.161.1034}.  The $np$ quasi-spin formalism was applied in Refs. \cite{tal93}.
For a system with two np pairs in a single-$j$ shell, it is natural  and very convenient to decompose the system into proton and neutron blocks. The wave function of a given state with total angular momentum $I$ can
be written as~\cite{Qi10},
\begin{equation}\label{4pn}
|\Psi_I\rangle=\sum_{J_{p}, J_n} X_I(J_pJ_n) | j_{\pi}^2(J_p)j_{\nu}^2(J_n);I \rangle,
\end{equation}
where $X_I(J_pJ_n)$ is the amplitude of the four-body wave function and $J_p$ and $J_n$ are even numbers denoting the angular momenta of the proton and neutron pairs, respectively. For example, in the hole-hole channel, the ground state wave function of $^{96}$Cd is calculated to be \cite{qi11},
\begin{eqnarray}
\nonumber|\Psi_0({\rm gs})\rangle&=& 0.76|[\pi^2(0)\nu^2(0)]_I\rangle + 0.57|[\pi^2(2)\nu^2(2)]_I\rangle \\
\nonumber &+&0.24 |[\pi^2(4)\nu^2(4)]_I\rangle+0.13|[\pi^2(6)\nu^2(6)]_I\rangle\\
 &+&0.14|[\pi^2(8)\nu^2(8)]_I\rangle.
\end{eqnarray}

The four nucleons can couple to spin $I=0$ to $2(2j-1)$ and isospin $T=0$, 1 and $2$. 
The single-$j$ Hamiltonian can be written as,
\begin{eqnarray}\label{4h}
\nonumber \langle  j_{\pi}^2(J_p)j_{\nu}^2(J_n);I |\hat{V}| j_{\pi}^2(J_p')j_{\nu}^2(J_n');I \rangle\\
= (V_{J_p}+V_{J_n})\delta_{J_pJ_p'}\delta_{J_nJ_n'}
+\sum_JM^I_J(J_pJ_n;J_p'J_n')V_J,
\end{eqnarray}
where the spin $J$ can take both even and odd values ($J=0$ to $2j$).
The symmetric matrix $M$ is given
 as
\begin{eqnarray}
\nonumber M^I_J(J_pJ_n;J_p'J_n')=\sum_{\lambda}4\hat{J}_p\hat{J}_n\hat{J}_p'\hat{J}_n'\hat{J}^2\hat{\lambda}^2\left\{
\begin{array}{ccc}
 J_p & J_n & I \\
 \lambda & j & j \\
\end{array}\right\}\\
\times\left\{\begin{array}{ccc}
 J_p' & J_n' & I \\
 \lambda & j & j \\
\end{array}
\right\}\left\{\begin{array}{ccc}
 j & j & J \\
 \lambda & j & J_n \\
\end{array}
\right\}\left\{\begin{array}{ccc}
 j & j & J \\
 \lambda & j & J_n' \\
\end{array}
\right\},
\end{eqnarray}
where $\hat{J}=\sqrt{2J+1}$ and $\lambda$ and $j$ are half integers. 
The number of nucleon pairs in the $n=4$ system can be calculated as,
\begin{eqnarray}
\nonumber C^I_J=|X_I(J_pJ_n)|^2(\delta_{J_pJ}+\delta_{J_nJ})\\
+\sum_{J_pJ_n;J_p'J_n'}X_I(J_pJ_n)M^I_J(J_pJ_n;J_p'J_n')X_I(J_p'J_n'),
\end{eqnarray}
where the first and second terms in the right-hand side give the numbers of identical nucleon pairs and proton-neutron pairs, respectively. 

The $12^+$ state in $^{52}$Fe and $I^{\pi}=16^+$ in the four-hole system of $^{96}$Cd below the double magic $^{100}$Sn have long been expected to be a spin-trap isomers. The latter case was measured recently \cite{PhysRevLett.107.172502}.  That means their energies are lower than those of the corresponding $10^+_1$ and $14^+_1$ states. In $0f_{7/2}$ shell, the correlation energy of the four-hole system  $12^+$ is,
\begin{equation}
E_{12}(^{52}{\rm Fe})=\frac{6}{13}\bar{V}_5+3\bar{V}_6+\frac{33}{13}\bar{V}_7,
\end{equation}
where $\bar{V}$ denotes two-hole matrix elements. The position of the $12^+$ state relative to the corresponding $10^+_1$ states is sensitive to the strength of interaction ${\bar V}_7$. The number of nucleon pairs for the 2 np system $16^+$ in the $j=9/2$ shell is,
\begin{equation}
E_{16}(^{96}{\rm Cd})=\frac{8}{17}\bar{V}_7+3\bar{V}_8+\frac{43}{17}\bar{V}_9.
\end{equation}
Again, the position of the $16^{+}$ state relative to the first $14^+$ state is sensitive to the strength of the aligned interaction matrix element $\bar{V}_9$.

\subsection{Spin-aligned np pair coupling}
To illustrate the idea of the spin-aligned pair mode we will start with the
simple example of a $2n$-$2p$ system within a single $j$ 
shell. 
One may re-express the wave function in Eq. (\ref{4pn}) in an \textit{equivalent} representation in terms of $np$ pairs. This can be done analytically with the help of the overlap matrix as
\begin{eqnarray}\label{over}
\nonumber\langle [\nu\pi(J_1)\nu\pi(J_2)]_I |[\pi^2(J_p)\nu^2(J_n)]_I\rangle \\
= \frac{-2}{\sqrt{N_{J_1J_2}}}\hat{J_1}\hat{J_2}
\hat{J_p}\hat{J_n}\left\{
\begin{array}{ccc}
j&j&J_p\\
j&j&J_n\\
J_1&J_2&I
\end{array}
\right\},
\end{eqnarray}
where $N$ denotes the normalization factor. The overlap matrix automatically 
takes into account the Pauli principle. With interactions taken from 
experimental data and Ref. \cite{Sch76}, we examined a few shells with a high 
degeneracy, i.e., $0f_{7/2}$, $0g_{9/2}$ and $0h_{11/2}$, values in the range
$X^2_{J_p=J_n=0}=0.51-0.62$ for the ground states of these even-even nuclei. This means that 
the normal isovector pairing coupling scheme $(\nu^2)_0 \otimes (\pi^2)_0$
accounts for only about half of the ground state wave functions.
Instead, we found that for these wave functions it is 
$X^2_{J_1=J_2=2j}=0.92-0.95$, i.e., they virtually
can be represented by the spin-aligned $np$ coupling scheme.
An even more striking feature is that the low-lying yrast states
are calculated to be approximately equally spaced and their spin-aligned $np$ 
structure is the same for all of them. Moreover,
the quadrupole transitions between these states
show a strong selectivity, since the decay to 
other structures beyond the $np$ pair coupling scheme is unfavored. It should be emphasized that only states with even angular momenta can be generated from the spin-aligned $np$ coupling for systems with two pairs. The maximum spin one can get is $2(2j-1)$. For a given even spin $I$, only one state can be uniquely specified from the coupling of two aligned $np$ pairs. The other states (and also states with odd spins or total isospin $T>0$) involve the breaking of the aligned pairs.

\begin{figure}
\hspace{0.4cm}\includegraphics[scale=0.37]{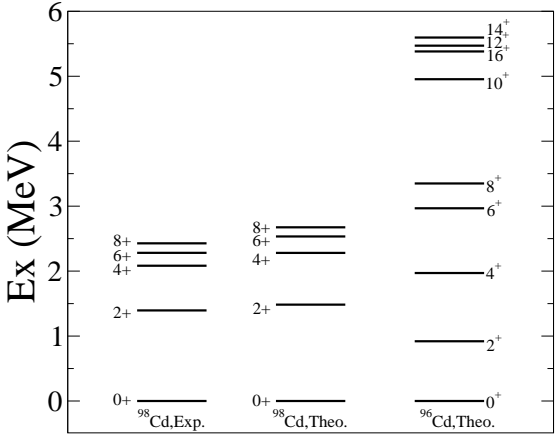}
\caption{Shell-model calculations for $^{96,98}$Cd in comparion with available experimental data. \label{96cd}}
\vspace{1cm}
\end{figure}

\begin{figure}
\includegraphics[scale=0.4]{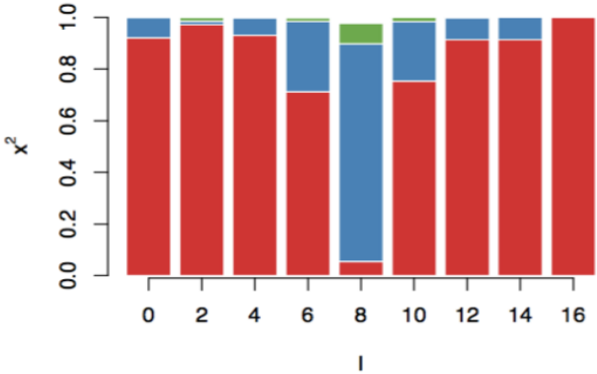}
\caption{Shell-model calculations for overlap between the full wave function and the spin-aligned np pair wave function for the first three states for each spin. \label{96cd2}}
\vspace{1cm}
\end{figure}

Shell-model calculations for the nuclei $^{96,98}$Cd are plotted in Fig. \ref{96cd}. In the former case, the yrast states are all found to be dominated by the spin-aligned np coupling except the $8^+_1$ state.  In that case the normal seniority coupling is favored in relation to the low energy of the isovector $8+$ pair in $^{98}$Cd. On the other hand, it is the second $T=0$ $8^+$ state in $^{96}$Cd that favors the spin-aligned np coupling. The overlap between the full wave function and the spin-aligned np pair wave funciton is given in Fig. \ref{96cd2}.

\subsection{The over-complete basis}
Calculations in Ref. \cite{qi11} were done within the standard shell-model framework with the help of two-body coefficients of fractional parentage. In the following we go through the formalism as used in Ref. \cite{Xu2012} within the so-called Multistep shell model approach (MSM), where the nn, pp and np pairs are considered on the same footing. 
We will use the Greek letter $\gamma_n$ to label the $n$-particle 
$np$ states. The $np$ states will be
$|\gamma_2\rangle=P^+(\gamma_2)|0\rangle$ where the $np$ creation operator is
$P^+(\gamma_2)= \sum_{i,p}X(ip;\gamma_2)c^+_ic^+_p$ and $c^+_i$ ($c^+_p$)
is the neutron (proton) single-particle creation operator. In the same fashion the
two-proton (two-neutron) creation operator will be denoted as $P^+(\alpha_2)$
($P^+(\beta_2)$). The four-particle
state, $|\gamma_4\rangle=P^+(\gamma_4)\vert 0\rangle$, is
\begin{multline}
\label{eq:wf4p}
P^+(\gamma_4)=\sum_{\alpha_2,\beta_2}X(\alpha_2\beta_2;\gamma_4)
P^+(\alpha_2) P^+(\beta_2)+\\
\sum_{\gamma_2\leq \gamma_2'}X(\gamma_2\gamma_2';\gamma_4)
P^+(\gamma_2)P^+(\gamma_2'),~~~
\end{multline}
where all possible like-particle and $np$ pairs are taken into account. In the two-pair case 
the basis elements $(\nu\nu)\otimes(\pi\pi)$ and $(\nu\pi)\otimes(\nu\pi)$ may 
be proportional to each other. The 
over-counting thus occurring is a result of describing the $np$ and 
like-particle excitations at the same time. 
Since the number of
MSM basis vectors is larger than the dimension of the shell model space, the wave function amplitudes $X$ are not well defined in our case
and, therefore,
they are not meaningful physically. The meaningful quantities
are the projections of the basis
vectors upon the physical vector, which we denote as \cite{lio81}
\bea
\label{eq:proj}
F(\alpha_2\beta_2;\gamma_4)&=&\langle\gamma_4|P^+(\alpha_2) P^+(\beta_2)|0\rangle,
\nn 
F(\gamma_2\gamma_2';\gamma_4)&=&\langle\gamma_4|P^+(\gamma_2)P^+(\gamma_2')|0\rangle.
\eea
The orthonormality condition now reads
\begin{multline}
\label{eq:ort4}
\delta_{\gamma_4\gamma_4'}=
\sum_{\alpha_2,\beta_2}X(\alpha_2\beta_2;\gamma_4)
F(\alpha_2\beta_2;\gamma_4') \\
+
\sum_{\gamma_2\leq \gamma_2'}X(\gamma_2\gamma_2';\gamma_4)
F(\gamma_2\gamma_2';\gamma_4).
\end{multline}

The norm of the MSM basis
$|\gamma_2\gamma_2'\rangle$ $=$ $P^+(\gamma_2)P^+(\gamma_2')|0\rangle$, i.e.,
$N(\gamma_2\gamma_2';\gamma_4)=
\sqrt{\langle\gamma_2\gamma_2'|\gamma_2\gamma_2'\rangle}$,
may not be unity. Therefore
the interesting quantity is not the projection $F$ but rather the cosine
of the angle between the basis vector and the physical vector, i.e.,  $\cos(\phi)=x$ and 
\be
\label{cos4}
x(\gamma_2\gamma_2';\gamma_4)=
F(\gamma_2\gamma_2';\gamma_4)/N(\gamma_2\gamma_2';\gamma_4).
\ee 
If we would have taken as basis elements the complete set of
orthonormal states
$\{P^+(\alpha_2) P^+(\beta_2)|0\rangle\}$ (which is the standard shell model
basis as used in Ref. \cite{qi11}) then the second term in Eq. (\ref{eq:ort4}) would not have appeared
and one would have obtained
$X(\alpha_2\beta_2;\gamma_4)=x^*(\alpha_2\beta_2;\gamma_4)$, as expected
in an orthonormal basis.
One thus sees that the advantage of the MSM basis is that one can extract
the physical structure of the calculated states just by examining the
quantity $x$.

The dynamic matrix of the two-neutron two-proton system is given as
\begin{multline}
    (W(\gamma_4)-W(\gamma_2)-W(\gamma'_2))\langle\gamma_4|(P^\dag(\gamma_2)P^\dag(\gamma'_2))_{\gamma_4}|0\rangle=\\
    \sum_{\gamma''_2\leqslant\gamma'''_2}\Bigg\{\sum_{p_1p_2n_1n_2}(-1)\frac{W(\gamma''_2)+W(\gamma'''_2)-\varepsilon_{p_1}-\varepsilon_{p_2}-\varepsilon_{n_1}-\varepsilon_{n_2}}{1+\delta_{\gamma''_2\gamma'''_2}}\\
\times(\mathbb{A}_1+\mathbb{A}_2)\Bigg\}\langle\gamma_4|(P^\dag(\gamma''_2)P^\dag(\gamma'''_2))_{\gamma_4}|0\rangle\\[0.5em]
    +\sum_{\alpha_2\beta_2}\Bigg\{\sum_{p_1p_2n_1n_2}\Big(W(\alpha_2)+W(\beta_2)-\varepsilon_{p_1}-\varepsilon_{p_2}-\varepsilon_{n_1}-\varepsilon_{n_2}\Big)\times \mathbb{B}\Bigg\}
    \\\langle\gamma_4|(P^\dag(\alpha_2)P^\dag(\beta_2))_{\gamma_4}|0\rangle,
\end{multline}
and
\begin{multline}
    (W(\gamma_4)-W(\alpha_2)-W(\beta_2))\langle\gamma_4|(P^\dag(\alpha_2)P^\dag(\beta_2))_{\gamma_4}|0\rangle=\\
    \sum_{\gamma''_2\leqslant\gamma'''_2}\Bigg\{\sum_{p_1p_2n_1n_2}\frac{W(\gamma''_2)+W(\gamma'''_2)-\varepsilon_{p_1}-\varepsilon_{p_2}-\varepsilon_{n_1}-\varepsilon_{n_2}}{1+\delta_{\gamma''_2\gamma'''_2}}\mathbb{C}\Bigg\}\\
    \times\langle\gamma_4|(P^\dag(\gamma''_2)P^\dag(\gamma'''_2))_{\gamma_4}|0\rangle,
\end{multline}
where $W$ denotes the corresponding $n$-particle energy. To obtain above equation we  have assumed  that  the  four-particle  system  was  decomposed  into  two different  blocks in terms of $(\pi\pi)\otimes(\nu\nu)$ and  $(\nu\pi)\otimes(\nu\pi)$.

The $\mathbb{A}$, $\mathbb{B}$ and $\mathbb{C}$ matrix elements are defined as,
\begin{multline}
    \mathbb{A}_1=(-1)^{2p_1+n_1+n_2+\gamma'_2+\gamma'''_2}\hat{\gamma}_2\hat{\gamma}'_2\hat{\gamma}''_2\hat{\gamma}'''_2\\
        \times X(p_1n_1;\gamma_2)X(p_2n_2;\gamma'_2)X(p_1n_2;\gamma''_2)X(p_2n_1;\gamma'''_2)\\
        \times\left\{
          \begin{array}{ccc}
            p_1 & n_1 & \gamma_2 \\
            n_2 & p_2 & \gamma'_2 \\
            \gamma''_2 & \gamma'''_2 & \gamma_4 \\
          \end{array}
        \right\},
\end{multline}
\begin{multline}
    \mathbb{A}_2=(-1)^{2p_1+n_1+n_2+\gamma'_2+\gamma'''_2+\gamma_4}\hat{\gamma}_2\hat{\gamma}'_2\hat{\gamma}''_2\hat{\gamma}'''_2\\
        \times X(p_1n_1;\gamma_2)X(p_2n_2;\gamma'_2)X(p_2n_1;\gamma''_2)X(p_1n_2;\gamma'''_2)\\\times\left\{
          \begin{array}{ccc}
            p_1 & n_1 & \gamma_2 \\
            n_2 & p_2 & \gamma'_2 \\
            \gamma'''_2 & \gamma''_2 & \gamma_4 \\
          \end{array}
        \right\},
\end{multline}
\begin{multline}
    \mathbb{B}=\hat{\gamma}_2\hat{\gamma}'_2\hat{\alpha}_2\hat{\beta}_2X(p_1n_1;\gamma_2)X(p_2n_2;\gamma'_2)Y(p_1p_2;\alpha_2)\\
    \times Y(n_1n_2;\beta_2)
        \left\{
          \begin{array}{ccc}
            p_1 & n_1 & \gamma_2 \\
            p_2 & n_2 & \gamma'_2 \\
            \alpha_2 & \beta_2 & \gamma_4 \\
          \end{array}
        \right\},
\end{multline}
and
\begin{multline}
    \mathbb{C}=\hat{\alpha}_2\hat{\beta}_2\hat{\gamma}''_2\hat{\gamma}'''_2Y(p_1p_2;\alpha_2)Y(n_1n_2;\beta_2)\\
    \times X(p_1n_1;\gamma''_2)X(p_2n_2;\gamma'''_2)
        \left\{
          \begin{array}{ccc}
            p_1 & p_2 & \alpha_2 \\
            n_1 & n_2 & \beta_2 \\
            \gamma''_2 & \gamma'''_2 & \gamma_4 \\
          \end{array}
        \right\}.
\end{multline}
In  all  cases  we  use  the  same  symbols  to label  states  as  well  as  the  corresponding  angular  momenta. The coefficient $Y$ is related to $X$ by $Y(ij;\alpha_2)=(1+\delta_{ij})^{1/2}X(ij;\alpha_2)$.

The overlap matrix is defined as follows,
\begin{multline}
    \langle0|(P^\dag(\gamma_2)P^\dag(\gamma'_2))^\dag_{\gamma_4}(P^\dag(\gamma''_2)P^\dag(\gamma'''_2))_{\gamma_4}|0\rangle\\=\delta_{\gamma_2\gamma''_2}\delta_{\gamma'_2\gamma'''_2}+(-1)^{\gamma_2+\gamma'_2+\gamma_4}\delta_{\gamma_2\gamma'''_2}\delta_{\gamma'_2\gamma''_2}\\
    -\sum_{p_1p_2n_1n_2}(\mathbb{A}_1+\mathbb{A}_2)\\
    \langle0|(P^\dag(\gamma_2)P^\dag(\gamma'_2))^\dag_{\gamma_4}(P^\dag(\alpha_2)P^\dag(\beta_2))_{\gamma_4}|0\rangle=\sum_{p_1p_2n_1n_2}\mathbb{B}\\
    \langle0|(P^\dag(\alpha_2)P^\dag(\beta_2))^\dag_{\gamma_4}(P^\dag(\alpha'_2)P^\dag(\beta'_2))_{\gamma_4}|0\rangle=\delta_{\alpha_2\alpha'_2}\delta_{\beta_2\beta'_2},
\end{multline}
which correspond to the overlap between states of the forms $\langle \nu\pi \otimes \nu\pi |\nu\pi \otimes \nu\pi \rangle$, $\langle \nu\pi \otimes \nu\pi |\nu\nu \otimes \pi\pi \rangle$ and $\langle \nu\nu \otimes \pi\pi |\nu\nu \otimes \pi\pi \rangle$, respectively.

\subsection{Systems with more than two np pairs}
It is challenging to to describe the wave functions of systems with more than two np pairs in relation to the over-completeness of the pair wave function.
On top of that, the np pairs can couple to many different configurations due to their non-zero angular momentum.
For the six-particle case we use the MSM partition of two- times
four-particles, as it was done in Ref. \cite{lio81} for systems with six like 
particles. Thus the corresponding 
wave function will be $|\gamma_6\rangle=P^+(\gamma_6)|0\rangle$, where
\be
\label{eq:wf6p}
P^+(\gamma_6)=
\sum_{\gamma_2,\gamma_4}X(\gamma_2\gamma_4;\gamma_6)
P^+(\gamma_2)P^+(\gamma_4),
\ee
and
\begin{multline}
|\gamma_6\rangle=\sum_{\gamma_2\gamma_4}X(\gamma_2\gamma_4;\gamma_6)\langle \gamma_2\gamma_4|\gamma_6\rangle\\
\times\sum_{\alpha_2\beta_2}X(\alpha_2\beta_2;\gamma_4)\langle \alpha_2\beta_2|\gamma_4\rangle\sum_{p_1n_1}X(p_1n_1;\gamma_2)\langle p_1n_1|\gamma_2\rangle\\
\times\frac{1}{2}\sum_{p_2p_3}Y(p_2p_3;\alpha_2)\langle p_2p_3|\alpha_2\rangle\\
\times\frac{1}{2}\sum_{n_2n_3}Y(n_2n_3;\beta_2)\langle n_2n_3|\beta_2\rangle\; p^\dag_1n^\dag_1p^\dag_2p^\dag_3n^\dag_2n^\dag_3|0\rangle.
\end{multline}
As before, we will evaluate the projection of the basis vectors upon the physical
vectors, i.e., $F(\gamma_2\gamma_4;\gamma_6)$.
In this six-particle case one can also view the MSM basis elements as the direct
tensorial product of three pairs which takes the forms $\nu\pi\otimes\nu\pi\otimes\nu\pi$ and $\nu\pi\otimes\nu\nu\otimes\pi\pi$.

For the partition of one $np$ pair times the 4-particle system, the dynamic matrix is given as
\begin{multline}
  \nonumber   \big(W(\gamma_6)-W(\gamma_2)-W(\gamma_4)\big)\langle\gamma_6|(\gamma_2^\dag\gamma_4^\dag)_{\gamma_6}|0\rangle=\\
    \sum_{\gamma'_2\gamma'_4}\bigg\{\sum_{p_1n_1n_2n_3}\sum_{\alpha_2\beta_2\beta'_2\theta_4}\\
    \big(W(\gamma'_2)+W(\beta'_2)-\varepsilon_{p_1}-\varepsilon_{n_1}-\varepsilon_{n_2}-\varepsilon_{n_3}\big)\mathbb{A}_1\\
    +\sum_{p_1p_2p_3n_1}\sum_{\alpha_2\beta_2\alpha'_2\phi_4}\big(W(\gamma_2)+W(\alpha'_2)-\varepsilon_{p_1}-\varepsilon_{p_2}-\varepsilon_{p_3}-\varepsilon_{n_1}\big)\mathbb{A}_2\bigg\}\\
    \times\langle\gamma_6|({\gamma'_2}^\dag{\gamma'_4}^\dag)_{\gamma_6}|0\rangle.
\end{multline}
The overlap matrix of $2\times4$ block is given as
\begin{multline}
 \langle0|(\gamma_2^\dag\gamma_4^\dag)^\dag_{\gamma_6}({\gamma'_2}^\dag{\gamma'_4}^\dag)_{\gamma_6}|0\rangle \\=
 \delta_{\gamma_2\gamma'_2}\delta_{\gamma_4\gamma'_4}+\sum_{p_1n_1n_2n_3}\sum_{\alpha_2\beta_2\beta'_2\theta_4}\mathbb{A}_1+\sum_{p_1p_2p_3n_1}\sum_{\alpha_2\beta_2\alpha'_2\phi_4}\mathbb{A}_2\\
    +\sum_{p_1p_2p_3n_1n_2n_3}\sum_{\alpha_2\beta_2\alpha'_2\beta'_2\psi_2}\mathbb{B}.
\end{multline}
Definations for $\mathbb{A}_1$, $\mathbb{A}_2$ and $\mathbb{B}$ can be found in Ref. \cite{Xu2012}.
The transformation from the $2\times4$ block to the  $2\times2\times2$ block is given as
\begin{eqnarray*}
    \langle\gamma_6|(\gamma_2^\dag \alpha_2^\dag \beta_2^\dag)_{\gamma_6}|0\rangle&=&\sum_{\gamma_4}\langle\gamma_6|(\gamma_2^\dag\gamma_4^\dag)_{\gamma_6}|0\rangle \langle\gamma_4|(\alpha_2^\dag \beta_2^\dag)_{\gamma_4}|0\rangle,\\
    \langle\gamma_6|(\gamma_2^\dag {\gamma'_2}^\dag {\gamma''_2}^\dag)_{\gamma_6}|0\rangle&=&\sum_{\gamma_4}\langle\gamma_6|(\gamma_2^\dag\gamma_4^\dag)_{\gamma_6}|0\rangle \langle\gamma_4|({\gamma'_2}^\dag {\gamma''_2}^\dag)_{\gamma_4}|0\rangle.
\end{eqnarray*}

We will describe the eight-particle states as
$|\gamma_8\rangle$ = $P^+(\gamma_8)|0\rangle$, where $P^+(\gamma_8)$ =
$\sum_{\gamma_4\leq \gamma_4'}X(\gamma_4 \gamma_4';\gamma_8)
P^+(\gamma_4)P^+(\gamma_4')$. Proceeding as above we will also evaluate the
cosine of the angle between $|\gamma_8\rangle$ and all the possible four-pair
states that can be formed.
\begin{multline}
    |\gamma_8\rangle=\sum_{\gamma_4\gamma'_4}X(\gamma_4\gamma'_4;\gamma_8)\langle\gamma_4\gamma'_4|\gamma_8\rangle \sum_{\alpha_2\beta_2}X(\alpha_2\beta_2;\gamma_4)\langle \alpha_2\beta_2|\gamma_4\rangle\\
    \sum_{\alpha'_2\beta'_2}X(\alpha'_2\beta'_2;\gamma'_4)\langle \alpha'_2\beta'_2|\gamma'_4\rangle\\
        \times\frac{1}{2}\sum_{p_1p_2}Y(p_1p_2;\alpha_2)\langle p_1p_2|\alpha_2\rangle \times\frac{1}{2}\sum_{n_1n_2}Y(n_1n_2;\beta_2)\langle n_1n_2|\beta_2\rangle \\ \times\frac{1}{2}\sum_{p_3p_4}Y(p_3p_4;\alpha'_2)\langle p_3p_4|\alpha'_2\rangle\\
         \times\frac{1}{2}\sum_{n_3n_4}Y(n_3n_4;\beta'_2)\langle n_3n_4|\beta'_2\rangle p^\dag_1p^\dag_2n^\dag_1n^\dag_2p^\dag_3p^\dag_4n^\dag_3n^\dag_4|0\rangle.
\end{multline}
The dynamic matrix for the $4\times 4$ block is defined as
\begin{multline}
    \big(W(\gamma_8)-W(\gamma_4)-W(\gamma'_4)\big)\langle\gamma_8|(\gamma_4\gamma'_4)_{\gamma_8}|0\rangle\\
    =\sum_{\gamma''_4\leqslant\gamma'''_4}\frac{1}{1+\delta_{\gamma''_4\gamma'''_4}}\sum_{\alpha_2\beta_2\alpha'_2\beta'_2}\\
    \begin{split}
        \Bigg\{\Big(W(\gamma''_4)+W(\gamma'''_4)-W(\alpha_2)-W(\beta_2)-W(\alpha'_2)-W(\beta'_2)\Big)\\
        \times(\mathbb{A}_1+\mathbb{A}_2)\\
            +\sum_{\alpha''_2\alpha'''_2}\sum_{p_1p_2p_3p_4}\Big(W(\alpha''_2)+W(\alpha'''_2)-\varepsilon_{p_1}-\varepsilon_{p_2}-\varepsilon_{p_3}-\varepsilon_{p_4}\Big)\\
            \times(\mathbb{B}_1+\mathbb{B}_2)\\
            +\sum_{\beta''_2\beta'''_2}\sum_{n_1n_2n_3n_4}\Big(W(\beta''_2)+W(\beta'''_2)-\varepsilon_{n_1}-\varepsilon_{n_2}-\varepsilon_{n_3}-\varepsilon_{n_4}\Big)\\
         \times   (\mathbb{C}_1+\mathbb{C}_2)\Bigg\}\langle\gamma_8|(\gamma''_4\gamma'''_4)_{\gamma_8}|0\rangle
    \end{split}
\end{multline}
The overlap matrix of $4\times4$ block is given by
\bea
 \nonumber   \langle0|(\gamma_4^\dag{\gamma'_4}^\dag)^\dag_{\alpha_8}({\gamma''_4}^\dag{\gamma'''}_4^\dag)_{\alpha_8}|0\rangle=\delta_{\gamma_4\gamma''_4}\delta_{\gamma'_4\gamma''_4}\\
 \nonumber +(-1)^{\gamma_4+\gamma'_4+\alpha_8}\delta_{\gamma_4\gamma'''_4}\delta_{\gamma'_4\gamma''_4}+\sum_{\alpha_2\beta_2\alpha'_2\beta'_2}(\mathbb{A}_1+\mathbb{A}_2)\\
\begin{split}
        +\sum_{\alpha_2\beta_2\alpha'_2\beta'_2}\sum_{\alpha''_2\alpha''_2}\sum_{p_1p_2p_3p_4}(\mathbb{B}_1+\mathbb{B}_2)\\
        \nonumber+\sum_{\alpha_2\beta_2\alpha'_2\beta'_2}\sum_{\beta''_2\beta'''_2}\sum_{n_1n_2n_3n_4}(\mathbb{C}_1+\mathbb{C}_2)\\
        +\sum_{\alpha_2\beta_2\alpha'_2\beta'_2}\sum_{\alpha''_2\beta''_2\alpha'''_2\beta'''_2}\sum_{p_1p_2p_3p_4}\sum_{n_1n_2n_3n_4}\mathbb{D},
\end{split}
\eea

Definitions for $\mathbb{A}$, $\mathbb{B}$ and $\mathbb{C}$ have been given in Ref. \cite{Xu2012}.The transformation from the $4\times4$ block to the $2\times2\times2\times2$ block is defined as,
\begin{multline}
\langle\gamma_8|(\alpha_2\beta_2\alpha'_2\beta'_2)_{\gamma_8}|0\rangle=\sum_{\gamma_4\gamma'_4}\langle\gamma_8|(\gamma_4\gamma'_4)_{\gamma_8}|0\rangle\\
\times\langle\gamma_4|(\alpha_2\beta_2)_{\gamma_4}|0\rangle\langle\gamma'_4|(\alpha'_2\beta'_2)_{\gamma'_4}|0\rangle,\\
  \langle\gamma_8|(\alpha_2\beta_2\gamma_2\gamma'_2)_{\gamma_8}|0\rangle=\sum_{\gamma_4\gamma'_4}\langle\gamma_8|(\gamma_4\gamma'_4)_{\gamma_8}|0\rangle\\
  \times\langle\gamma_4|(\alpha_2\beta_2)_{\gamma_4}|0\rangle\langle\gamma'_4|(\gamma_2\gamma'_2)_{\gamma'_4}|0\rangle,\\
    \langle\gamma_8|(\gamma_2\gamma'_2\gamma''_2\gamma'''_2)_{\gamma_8}|0\rangle=\sum_{\gamma_4\gamma'_4}\langle\gamma_8|(\gamma_4\gamma'_4)_{\gamma_8}|0\rangle\\
    \times\langle\gamma_4|(\gamma_2\gamma'_2)_{\gamma_4}|0\rangle\langle\gamma'_4|(\gamma''_2\gamma'''_2)_{\gamma'_4}|0\rangle.
\end{multline}

In Ref. \cite{Xu2012} we applied the method to study the spin-aligned $np$ pair coupling scheme 
\cite{ced11,qi11}. We restricted restrict our calculations to the single $0g_{9/2}$ 
shell with the interaction matrix elements taken from Ref.~\cite{qi11}.
The formalism described above can be 
naturally generalized to systems with many shells. However, the computation can be very heavy. A parallel algorithm is underdevelopment.
\subsection{Spin-aligned coupling and quarter coupling}

\begin{table}
\begin{center}
\caption{Configurations with the largest probabilities for the state $^{92}$Pd($0^+_1$)
corresponding to the tensorial products of different two-particle
states (upper) and four-particle states (lower). From Ref. \cite{Qi2012b}.}
\label{fbos0}
\vskip2mm
\begin{tabular}{c|c}
\hline
Configuration   & $x^2$ \cr
\hline
$|\gamma_2=9^+\gamma_2'=9^+\gamma_2''=9^+\gamma_2'''=9^+\rangle$& 0.85  \cr
$|\gamma_2=9^+\gamma_2'=9^+\alpha_2=0^+\beta_2=0^+\rangle$& 0.76  \cr
$|\gamma_2=8^+\gamma_2'=1^+\alpha_2=0^+\beta_2=8^+\rangle$& 0.56  \cr
$|\gamma_2=8^+\gamma_2'=1^+\alpha_2=8^+\beta_2=0^+\rangle$& 0.56  \cr
$|\gamma_2=1^+\gamma_2'=1^+\alpha_2=0^+\beta_2=0^+\rangle$& 0.52  \cr
\hline
$|\gamma_4=0^+_1\gamma_4'=0^+_1\rangle$&  0.98  \cr
$|\gamma_4=8^+_1\gamma_4'=8^+_1\rangle$&  0.94  \cr
$|\gamma_4=8^+_2\gamma_4'=8^+_2\rangle$&  0.92  \cr
$|\gamma_4=16^+_1\gamma_4'=16^+_1\rangle$&  0.81  \cr
\hline
\end{tabular}
\end{center}
\end{table}

As mentioned, the four $J=9$ $np$ pairs in $^{92}$Pd can couple in various ways. In Ref. \cite{qi11} it was found with the help of two-particle fractional parentage that the dominating components can be well represented by a single configuration $((((\nu\pi)_9\otimes(\nu\pi)_9)_{I'=16}\otimes(\nu\pi)_9)_{I''=9}\otimes(\nu\pi)_9)_{I}$. 
Now we are able to re-project the wave function on the different coupling of np pairs. For the system with four np pairs we can write the wave functions as the coupling of four independent pairs, three np pair coupled to one pair as above, as well as the coupling of two four-particle states.
As mentioned in Ref. \cite{Xu2012}, the MSM basis is highly over-complete. we calculated in Table \ref{fbos0} the quantities $x$, i.e., the cosines 
of the angles between the vectors 
$|\gamma_8\rangle$ and all the possible vectors  that can be formed by the coupling of
four pairs for the ground state of $^{92}$Pd. Since many
combinations are similar to each other there is not a value of $x$ which is significantly
larger than the others. But the most important MSM configuration
is the one corresponding to
the four $9^+$ aligned pairs. The second one is a combination of two
aligned $9^+$ states and the normal pairing states. For the normal pairing state it is
$x^2(\alpha_2=0^+,\beta_2=0^+\alpha_2'=0^+\beta_2'=0^+;\gamma_8=0^+_1)$
= 0.46.
 
In the analysis in the lower panel of Table \ref{fbos0} for eight-particle systems like $^{92}$Pd, we choose as basis the partition 
$|\gamma_4\gamma_4';\gamma_8\rangle$. 
But one finds, again, that
for the ground state of $^{92}$Pd the most important MSM configuration
is the one corresponding to $|\gamma_4=0^+_1\gamma_4'=0^+_1\rangle$, that is, the coupling of two states each being composed of
two $9^+$ aligned pairs. This correspond to some kind of four-body quartet coupling.
Bearing this in mind, it may be interesting to compare present calculations with the stretch scheme proposed in Ref.~\cite{PhysRev.161.1034}. In that
scheme the dynamics of a $2N$ system is determined by the coupling of two 
$(\nu\pi)^{N/2}$ maximally aligned (stretch) vectors. For the $4n-4p$ system of $^{92}$Pd, the corresponding  wave function for a state with total angular momentum $I$ can be written as $((\nu\pi)^2_{I_1=16}\otimes(\nu\pi)^2_{I_1=16})_{I}$ where $(\nu\pi)^2_{I_1=16}$ denotes the stretch vector. It corresponds to the unique configuration of  $((\nu\pi)_9\otimes(\nu\pi)_9)_{I_1=16}$ in the $np$ pair coupling scheme. 
The configuration that has the second largest projection corresponds to $|\gamma_4=8^+_1\gamma_4'=8^+_1\rangle$. This is an interesting phenomenon that deserves further investigation. As seen in Fig. \ref{96cd2}, the $|\gamma_4=8^+_1\rangle$ state in $^{96}$Cd actually is of seniority-like type that is composed of two $J=0$ and 8 pairs. Other $|\gamma_4\gamma_4';\gamma_8\rangle$ configurations not shown in the table have more smaller overlap with the total wave function.

\subsection{Mixing between different orbitals}

In order to explore the importance of configuration mixing from other shells in determining the structure of $N=Z$ nuclei of concern here, we have performed shell model calculations in a variety of model spaces with Hamiltonians from Ref. \cite{hon09} and references quoted therein. We thus noticed that these
calculations provide practically the same
results for most properties of the low-lying yrast states in nuclei just below $^{100}$Sn, including the spectra and E2 decay properties. 
The quadrupole moment can be more sensitive to the mixing to other shells, for which the simple $0g_{9/2}$ shell calculation may not be enough.

It is beyond the scope to cover that area but it may be interesting to mention that it is still a challenging and tricky issue on how to understand
the effective shell-model wave functions and the mixing between different components, which in principle are model dependent quantities and have a non-observable nature
\cite{Talmi2009}. There are quite extensive studies on this issue now from different perspective including the non-observability of spectroscopic factors.

\subsection{The $0f_{7/2}$-shell nuclei}
\begin{figure}
\includegraphics[scale=0.33]{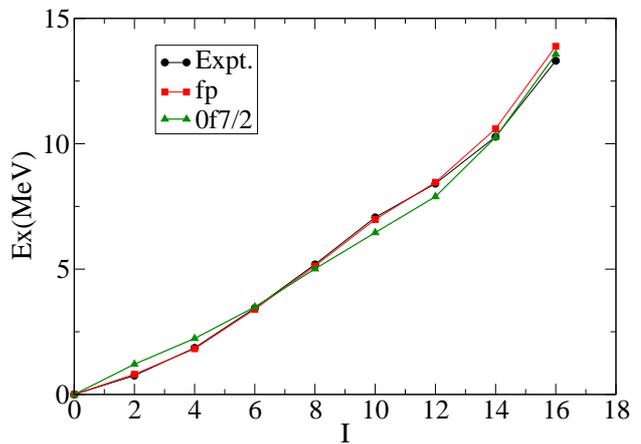}
\caption{Calculations on the energies of the yrast states in $^{48}$Cr in the full $fp$ and $0f_{7/2}$ shell model spaces in comparison with experimental data. \label{cr48e2}}
\end{figure}

\begin{figure}
\includegraphics[scale=0.33]{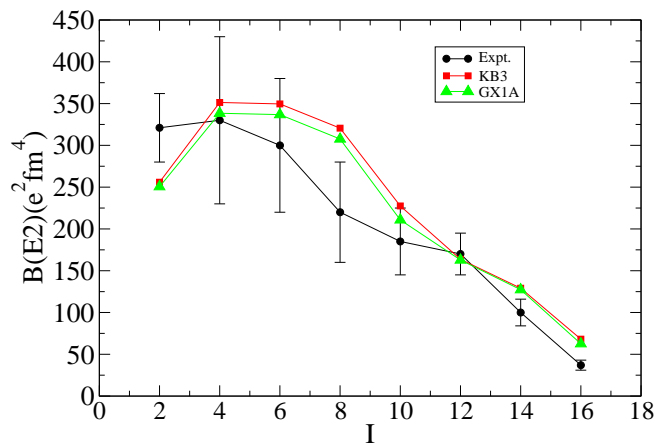}
\caption{Comparion between experimental B(E2) values of $^{48}$Cr along the yrast line and calculations with different interactions.\label{crbe2}}
\end{figure}

One interesting question is that why the spin-aligned np pair coupling is not observed in lighter $N=Z$ nuclei, in particular those in the $0f_{7/2}$ orbital between $N=Z=20$ and 28. Actually, as can be seen from Fig. \ref{cr48e2}, calculations in the single-$0f_{7/2}$ shell indeed predict a rather equally-spaced pattern for the yrast band upto around 
$I=12$. A closer look at the wave function show that, in that case, the wave function is indeed dominated by the spin-aligned $7^+$ np pair coupling. On the other hand, it is only when the $1p_{3/2}$ is included that one can reproduce the observed rotational like spectrum. It indicates that an essential physics is missing in the single-$j$ shell calculation, which, as we understand now, correspond to the quadrupole-quadrupole correlation between the shells $f_{7/2}$ and $p_{3/2}$. That coupling induce quadrupole deformation for which the Nilsson scheme is favored. The B(E2) values for the transitions around the yrast line in $^{48}$Cr is plotted in Fig. \ref{crbe2}  and compared with those given by shell-model calculations. The possible onset of $T=0$ pairing and deformations in high spin states of the $N=Z$ nucleus $^{48}$Cr is studied in Ref.\cite{Terasaki1998}.

\begin{figure}
\includegraphics[scale=0.43]{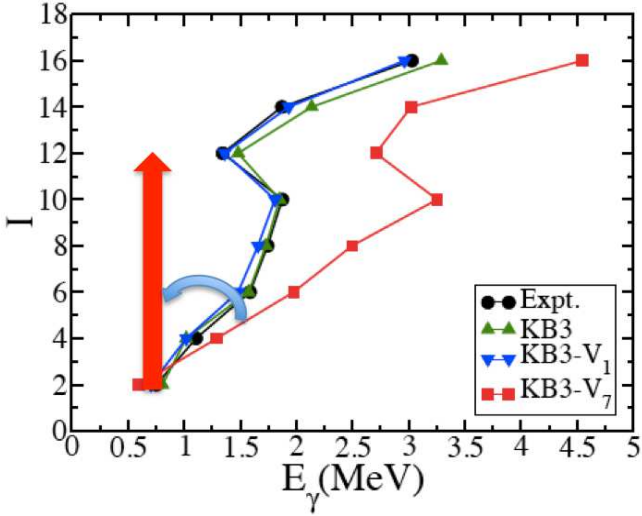}
\caption{\label{cr48}Illustration of the role played by $J=1$ and $J=7$ interaction matrix elements of the $f_{7/2}$ shell on the spectrum of $^{48}$Cr. The green, blue and red lines correspond to calculations with the full shell-model effective interaction and those with the $J=1$ np matrix element removed and with the $J=7$ np matrix element removed, respectively. A equally-spaced spectrum is expected (red arrow) if the $J=7$ np matrix is made stronger.}
\end{figure}

A simple but illustrating way to show the importance of different pairs is to calculate their dynamic effects on the spectrum.
In Fig. \ref{cr48} we did two calculations by removing the $J=1$ and $J=7$ interaction matrix elements of the $f_{7/2}$ orbital separately.
It can be seen clearly that the aligned $J=7$ np pair has a significant influence on the spectrum. A vibrational-like spectrum is indeed expected if one enhance that matrix element. On the other hand, the anti-aligned $J=1$ pair plays a much less significant role, which indicate that the $L=0$ pairs only take into account a minor fraction of the total np pairing.
 
\begin{figure}[htdp]
\includegraphics[scale=0.37]{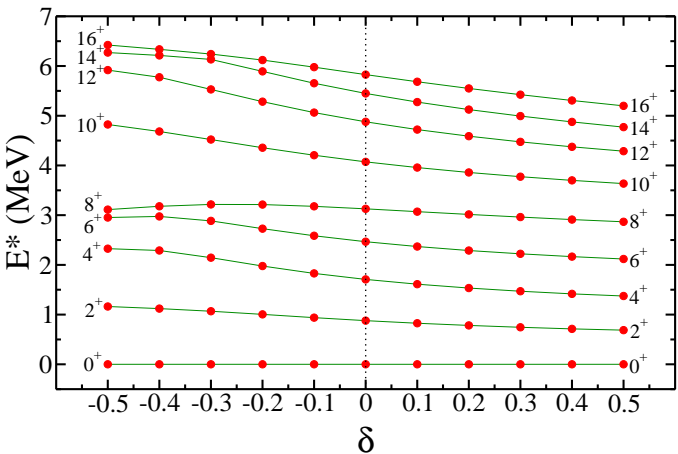}\\
\caption{(color online).
Shell model spectrum of $^{92}$Pd  as a function of a controlling parameter $\delta$ added by taking as interaction matrix element
the value $\mathcal{V}_9(\delta)= V_9(1+\delta)$. From Ref. \cite{qi11}.}
\label{vdel}
\end{figure}

As a comparison, 
in Fig. \ref{vdel} we present the resulting energy levels in $^{92}$Pd by scaling the strength of the $J=9$ pair interaction matrix element as $\mathcal{V}_9(\delta)= V_9(1+\delta)$. 
The figure clearly reveals that an increasing $\mathcal{V}_9(\delta)$ results in a vibrational like spectrum. In contrast, as 
$\mathcal{V}_9(\delta)\rightarrow 0$ a seniority-like situation is reached. This coincides
with the analysis of the $np$ interaction above, showing that the pair mode $(g_{9/2}^2)_9$
dominates the spectrum. Again, contributions from other coupled pairs to the spectral structure are much less pronounced.  Calculations in different model spaces show a similar pattern.

\section{The np pairing correlation within the deformed shell-model framework}
\label{Sect.01}

In this work, we concentrated on the np correlation within the spherical shell-model framework. It should be mentioned that there has been a long history and extensive studies with the Nilsson scheme plus np correlations of the BCS type as well as exact pairing calculations. Unlike the shell model, the simpler BCS and HFB approaches provide a physically much more clear tool to study pairing correlation in atomic nuclei, even though the generalization of BCS and HFB techniques to incorporate the interplay of $T=0$ and $T=1$ pairs on equal footing is by itself non-trivial.  It is beyond the scope of the present work to give a full picture of studies in that direction. A detailed review can be found in Ref. \cite{Frauendorf201424}. 
Here we mentioned a few studies around the same region as we studied above. 

$N\sim Z$ nuclei around $^{80}$Zr are expected to exhibit rich nuclear shape effects in relation to the QQ correlation between $g_{9/2}$ and $d_{5/2}$ orbitals and the intrusion of $h_{11/2}$.  A possible transition from isovector to isoscalar pairing condensate at high angular momenta may be expected.  The $T=0$ np pairing may also play an important role in the prediction of the super-deformed bands in nuclei around $^{60}$Zr and $^{88}$Ru.
There is ongoing effort trying to extend the spectroscopy of medium-spin states in $^{88}$Ru and neighboring nuclei in relation to the existence of super-deformed minimum in nuclei in this region, which may provide further evidence for the elusive isoscalar np pairing mode
\cite{Cederwall2015}. The idea behind is that, in high-spin rotational states of $N=Z$ nuclei, the $T=0$ pairing correlations can still be active.  Isovector pairing is expected to be suppressed by the Coriolis anti-pairing effect. In In Ref.
\cite{Back1999} super-deformed rotational bands observed in  the nucleus $^{88}$Mo. 
The low-spin states in this nucleus was studied recently in Ref. \citep{Andgren2007}.

Cranked Strutinsky-Woods-Saxon calculations based on a doubly stretched QQ np force were done in Ref. \cite{Satula1993573} to study the band crossings in intruder configurations of odd-$A$ nuclei.
Cranked shell model calculations with a zero-range residual np interaction were done in Ref. 
\cite{PhysRevC.59.1400} to study the rotational alignment in  $N\sim Z$ nuclei. It was shown  
 that the alignment
of high-$j$ nucleons with the rotational axis is sensitive
to the np interaction. The alignment of one kind of particles can be delayed
if the other kind of particles is present in the same
$j$ shell.

Hartree-Fock-Bogoliubov calculations for the $N=Z$ nucleus $^{80}$Zr was carried out in Ref. \cite{Goodman2001}, which give a ground state band with $T=1$ Cooper pairs and an excited band with $T=0$ Cooper pairs. It was found that, for the $T=0$ pair band, the dominant angular momentum for the pairs is $J=5$ instead of the aligned $J=9$. A systematic calculation for nuclei in this region was also done in Ref. \cite{Goodman1999}.
It indicates that there could be a transition in $N=Z$ nuclei from $T=1$ np-pairing to a predominantly $T=0$ pairing mode above $A=90$, with the intermediate mass 80-90 region showing a co-existence of T=0 and T=1 pairing modes.
A further enhancement in heavier $N=Z$ nuclei may be expected \cite{PhysRevLett.106.252502}.

In Ref.\cite{Delion2010}
the np correlations in  in Te and Xe isotopes above the Z = 50 shell closure are investigated . It is noticed that  the behavior of the $2^+$ and $4^+$ states in Te and Xe isotopes, which remain at a rather constant energy as one approaches the shell closure at N = 50 \cite{Sandzelius2007}, cannot be reproduced by standard quasi-particle random phase approximation calculations. To reproduce the experimental data within this model, one has to include a variable np interaction.

A cranked mean-field model with both $T = 1$ and $T = 0$ pairing interactions is proposed in Ref. \cite{Satula1997}, which includes the simultaneous presence of both pairing modes. It was suggested that the additional binding energy due to the $T = 0$ np pairing can be a possible microscopic explanation of the Wigner energy term. The co-existence of $T = 1$ and $T = 0$ is also studied in Ref. \cite{Sheikh2000}. The effect of deformation on the co-existence is also studied recently in Ref. \cite{PhysRevC.91.014308}. The influence of the np pairing and deformation on the neutrinoless double-$\beta$ decay in $^{76}$Ge is calculated in Ref. \cite{PhysRevC.90.031301}.
The Band crossings in intruder configurations of odd-A nuclei is studied in Ref.\cite{Satula1993573}. 
 The response of pairing correlations to rotation in the so-called isospace is investigated in Ref.  \cite{Satula2001} It is seen that the isovector pairing rather modestly modifies the single-particle moment of inertia in the isospace. In Ref. \cite{Wyss2007} the different behavior in the rotational structure of  Kr-73 and Rb-75 is suggested to be a fingerprint for the $T=0$ pairing. The negative-parity band of the former nucleus can only be reproduced by considering  the $T=0$ pairing.

\begin{figure}[htdp]
\includegraphics[scale=0.4]{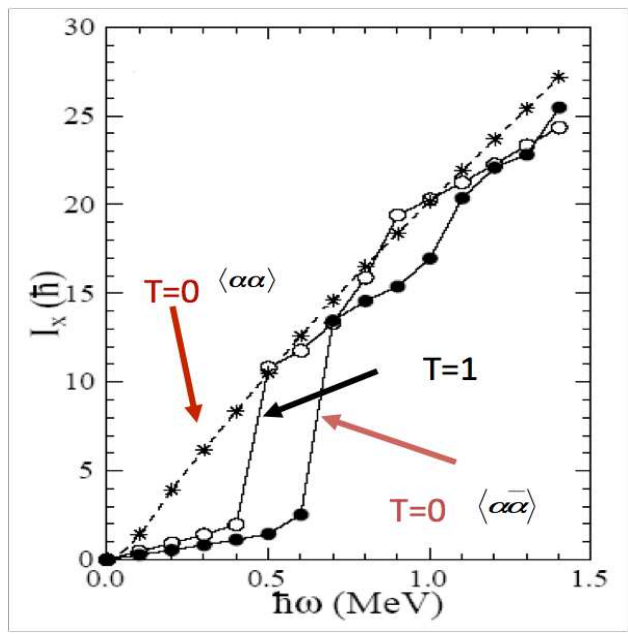}\\
\caption{(color online).
Response of the $T=1$ and $T=0$ pairing to the rotation. }
\label{wyss}
\end{figure}

As discussed in Ref. \cite{Goodman1999}, the $T=1$ pairing scatters pairs in opposite signature orbits. They are of the type $<\alpha\bar{\alpha}>$ where the bar indicates that the second nucleon
in a pair occupies a space-spin orbital that is the time reverse
of the first. In the meanwhile, the $T=0$ pair can be of both types $<\alpha\bar{\alpha}>$ and $<\alpha\alpha>$. They can scatter between orbitals of both opposite and same signatures. The second type corresponds to the aligned np pair. However,
as can be seen from Fig. \ref{wyss}, the $T=0$  $<\alpha\alpha>$  pairing exhibits a rotational like behavior as a function of frequency. This is in contrast with the shell-model calculations as discussed above. Further investigation in this direction is necessary.

A simple SO(5) seniority-like model was applied in Ref. \cite{Engel1996211} to study the interplay between like-particle and neutron-proton isovector pairing near N = Z. 
An algebraic description of the isovector and isoscalr pairing through the reductions of compact symplectic $Sp(4)$ 
symmetry was proposed in Ref. \cite{0954-3899-29-6-325}. The description works for both spherical and deformed systems.
The
SO(8) algebraic model has also been extensively applied in the study of the np pairing, in particular that in the $L=0$ channel \cite{Evans198177,PhysRevC.57.688,PhysRevC.63.034320}.

\section{$\alpha$ decay of $N\sim Z$ nuclei}

it might be interesting to pose a question whether the formation probabilities of the neutron-deficient isotopes with $N\sim Z$ are larger compared to their neutron rich counterparts. 
If it is indeed correct, that would mean that the cluster formation increases when protons and neutrons occupy the same shells \cite{Liddick2006,Seweryniak2006}. 
Refs. \cite{Liddick2006} compared the $\alpha$-decay reduced widths for Xe and Te nuclei with that of $^{212}$Po and neighboring Po isotopes and an enhancement by a factor of 2-3 is seen. 
We also noticed that  the $|RF(R)|^2$ value of $^{194}$Rn is larger by a similar factor  compared to the $|RF(R)|^2$ of the textbook $\alpha$-decay isotope $^{212}$Po \cite{Andreyev2013,Qi2014203}.
This faster  $\alpha$ decay
would change the borderline of accessible neutron deficient $\alpha$-decaying nuclei and might be a important question and motivation for further experimental work.

We go through very briefly the microscopic $R$-matrix description of the $\alpha$ decay. Details may be found in recent publications in Ref. \cite{Delion2010a,Astier2010a,Delion2006,Peltonen2008} and Refs. \cite{Qi2009,QI2009a,Qi2012c,Qi2010c}. The $\alpha$-decay half-life can be 
written as 
\begin{equation}\label{life}
T_{1/2}= \frac{\ln2}{\nu} \left|
\frac{H_l^+(\chi,\rho)}{RF_{\alpha}(R)} \right|^2,
\end{equation}
where $\nu$ is the velocity of the emitted $\alpha$ particle with angular momentum $l$.
$R$ is a distance chosen around the
nuclear surface where the internal wave function is matched with the outgoing cluster wave function. 
$H^+$ is the Coulomb-Hankel function with $\rho=\mu\nu R/\hbar$ and
$\chi = 4Ze^2/\hbar\nu$. $\mu$ is the reduced mass 
and $Z$ is the charge number of the daughter
nucleus. The quantity $F_{\alpha}(R)$ is the formation
amplitude of the $\alpha$ cluster at distance $R$.   The reduced width introduced in Ref. \cite{PhysRev.113.1593} is also a similar but effective quantity that depends on the effective optical potential.

The formation
amplitude $F(R)$ can be extracted from the experimental
half-lives by
\begin{equation}
\log |RF(R)|=\frac{1}{2}\log \left[ \frac{\ln
2}{\nu}|H^+_0(\chi,\rho)|^2\right] - \frac{1}{2}\log T^{{\rm
Expt.}}_{1/2}.
\end{equation}
This is done in Refs. \cite{Andreyev2013,Qi2014203} where a generic pattern for the systematics of the formation
amplitude $F(R)$ was also proposed. It was found that, when going from one isotope to another, the $\alpha$-particle formation probability usually varies much less than the penetrability. In other words, it is a consequence of the smooth variation in the nuclear structure that is often found when going from a nucleus to its neighbors. This is also the reason why, for example, the BCS approximation works so well in many regions of nuclei. 
In particular, within the BCS approach, it may be interesting to mention that the corresponding pairing gap is given by
$
\Delta=G\sum_i u_iv_i$,
where $G$ is the pairing strength, and $u_i$, $v_i$ are the standard occupation numbers. This implies that the pairing gaps can serve as a signature of the
change in  two-particle  correlation/clusterization, since they are also 
proportional to $\sum_k u_kv_k$. This feature is also responsible for the clustering 
of the four nucleons that eventually constitute the $\alpha$-particle at the 
nuclear surface of heavy nuclei.

\begin{figure}[htdp]
\includegraphics[scale=0.43]{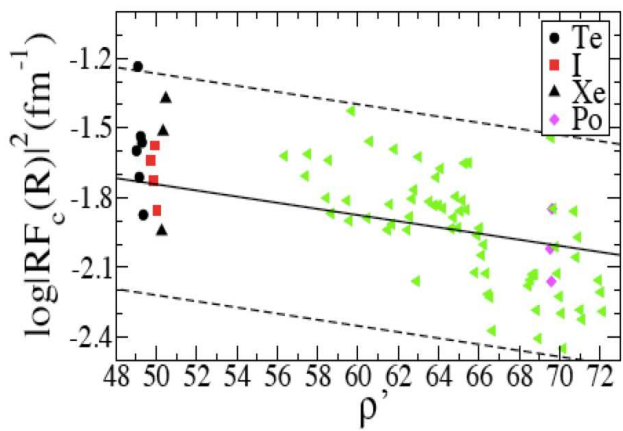}\\
\caption{(Color online) $\log_{10}|RF(R)|^2$ as a function
of $\rho'$ (see, Ref. \cite{Qi2009} for the definition) for nuclei around $^{100}$Sn in comparison with those of heavier isotopes. The solid line denotes the smooth behavior of the averged formation probability. The values between the two dashed lines differ from the corresponding
averge values by a factor of three.}\label{fvsrp}
\end{figure}

\begin{figure}[htdp]
\includegraphics[scale=0.4]{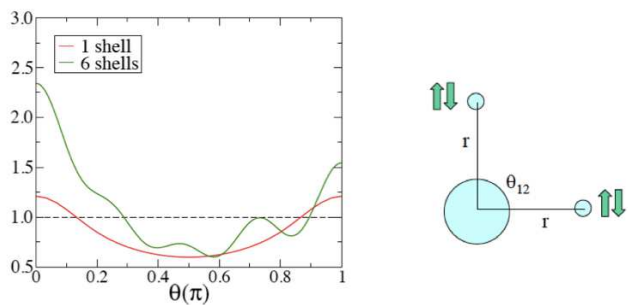}\\
\caption{(Color online) Schematic plot for the the influence of np correlation on the alpha formation amplitute in $^{104}$Te. $\theta$ is the angle is between the nn and pp pairs for which the two-particle clustering are induced mainly by the $T=1$ pairng. }\label{fvs}
\end{figure}

In Fig. \ref{fvsrp} we compared the $\alpha$ formation probabilities in nuclei just above $^{100}$Sn with those from the heavier nuclei. The formation certain low-lying show an increasing tread as the mass number decreases. This is in relation to the fact that the size of the nucleus also gets smaller, which favors the formation of $\alpha$ particles on the surface. In the meanwhile, the $\alpha$ formation probabilities in those lighter nuclei as shown in the figure follows the general trend but with a rather large fluctuations.

The $\alpha$ formation amplitude may increase as a result of enhanced $T=1$ nn and pp pairing and np correlation. In Fig. \ref{fvs} we illustrated the role played by the np correlation on the $\alpha$ formation amplitude. In reality, to evaluate the four-body correlation and its overlap with the alpha particle, one also decompose the four-particle wave function as the coupling of the proton pair and neutron pair. The two-body clustering of the proton and neutron pair is mainly induced by the nn and pp correlation. 
There is no correlation between the last neutron and proton pairs in our calculations for heavy nuclei \cite{Qi2010c} in relation to the neglect of np correlation. This is reasonable since the low-lying neutron and proton single-particle states are very different from each other in those cases and the np correlation is weak.
In Fig. \ref{fvs} we first evaluated the nn and pp two-body clustering in $^{102}$Sn and $^{102}$Te and then evaluated the correlation angle between the two pair by switching on and off the np correlation.
$\theta$ shows a uniform distribution if no np correlation is considered. If the np correlation is switched on, in particular if a large number of levels is included, there is significant enhancement of the four-body clustering at zero angle. This is eventually proportional to the $\alpha$ formation. It should be mentioned that, one need large number of orbitals already in heavy nuclei in order to reproduce properly the $\alpha$ clustering at the surface. The inclusion of np correlation will make the problem even more challenging due to the huge dimension. Work in this direction is under way.

\section{Summary and discussions}
In this contribution we discussed some studies on the elusive isoscalar and isovector np pairing modes. The basic idea behind is that, 
for nuclei with
$N\approx Z$,  the protons and neutrons near the Fermi surface occupy identical orbitals.  The $np$ pairs thus formed can couple to
isospin $T=1$ (isovector) or $T=0$ (isoscalar). It is known that, for a short range interaction, 
 the favored angular momenta are  $J=0$ for $T=1$ pairs and $J=1$ or $J=J_{max}$ for $T=0$ pairs.
Isospin symmetry and the   charge independence of the nuclear force implies
that for $N=Z$ nuclei, $J=0,~T=1$ $np$ pairing should exist on an equal
footing with $J=0,~T=1$ $nn$ and $pp$ pairing, which, however, it is an open question. Another interesting question relates to  the consequences of the strong attraction between the proton and neutron with $J=1$ and $J=J_{max}, ~T=0$. Despite vigorous activity over the last decade or so, the fundamental questions concerning the basic building blocks and fingerprints of above np pairing modes are still a matter of considerable debate, even though it is rather commonly believed that the $J=1, T=0$ pairing will not influence the low-lying spectroscopy in a major way. The nuclear shell model is one of the most accurate approaches in studying above effects and other properties of the low-lying states in atomic nuclei and the study of np pair correlation can provide a strict test to the specific parts of the effective interaction. It can also provide reliable prediction on the possible evidence of np interaction in certain low-lying as well as isomeric states. The challenging task is how to understand the pair content of the shell model wave function.

The spin-aligned np pair coupled scheme was introduced recently in relation to the observation of a vibrational-like spectrum in $N=Z$ nuclei $^{92}$Pd. A similar scheme is also expected in the heavier $^{96}$Cd. Both nuclei are expected to be spherical or weakly deformed and its main properties can be well described by the coupling of valence particles within the $g_{9/2}$ shell.  On the other hand, nuclear rotation in the presence of aligned $T=0$ pairing correlations resembles classical rigid body like rotation.

There are extensive efforts from both experimental and theoretical sides in studying the structure of $N\sim Z$ nuclei around $^{100}$Sn. It is hoped that those studies, in particular the measurement of nuclear masses, spectroscopy, reaction as well as $\alpha$ decay, will hold important clues to the np coupling schemes and the nature of $T=0$ pairing. 

\begin{ack}
This work was supported by the Swedish Research Council (VR) under grant Nos. 621-2012-3805, 621-2013-4323. We also thank the Swedish National Infrastructure for Computing (SNIC) at NSC in Link\"oping and PDC at KTH, Stockholm for computational support.
\end{ack}

\providecommand{\newblock}{}

\end{document}